\documentclass[onecolumn,aps,prd,showpacs,superscriptaddress,nofootinbib,amsmath,amssymb,floats,floatfix,showkeys,notitlepage,longbibliography]{revtex4-2}

\usepackage{graphicx}
\usepackage[dvipsnames]{xcolor}
\usepackage{subfigure}
\usepackage{palatino}
\usepackage[commandnameprefix=always]{changes}
\usepackage{hyperref}
\hypersetup{colorlinks=true,linkcolor=blue,urlcolor=blue,citecolor=blue}
\usepackage[toc,page]{appendix}
\usepackage[normalem]{ulem}

\usepackage{orcidlink}
\usepackage{lipsum}
\usepackage{graphicx}
\usepackage{subfigure}
\usepackage{palatino}
\usepackage{sans}
\usepackage{adjustbox}
\usepackage{latexsym}
\usepackage{amsmath}
\usepackage{amssymb}
\usepackage{amsfonts}
\usepackage{dcolumn}
\usepackage{bm}
\usepackage{tikz}
\usepackage{bigints}
\usepackage{array,tabularx,multirow,booktabs}
\usepackage[tracking=true]{microtype}
\SetTracking{}{500}
\SetTracking{encoding={*}, shape=sc}{40}
\UseRawInputEncoding %for inputenc error%
\allowdisplaybreaks
\usepackage{adjustbox}
\usepackage{latexsym}
\usepackage{amsmath}
\usepackage{amssymb}
\usepackage{amsfonts}
\usepackage{dcolumn}
\usepackage{bm}
\usepackage{tikz}
\usepackage{bigints}
\usepackage{array,tabularx,multirow,booktabs}
\usepackage[tracking=true]{microtype}
\usepackage{color}
\UseRawInputEncoding %for inputenc error%
\allowdisplaybreaks

\def\0{{\sst{(0)}}}
\def\1{{\sst{(1)}}}
\def\2{{\sst{(2)}}}
\def\3{{\sst{(3)}}}
\def\4{{\sst{(4)}}}
\def\5{{\sst{(5)}}}
\def\6{{\sst{(6)}}}
\def\7{{\sst{(7)}}}
\def\8{{\sst{(8)}}}
\def\sst#1{{\scriptscriptstyle #1}}

\def\ii{{\rm i}}

\newcommand\beq{\begin{equation}}
\newcommand\eeq{\end{equation}}
\newcommand\beqn{\begin{eqnarray}}
\newcommand\eeqn{\end{eqnarray}}

\begin{document} \sloppy

\title{Black hole with global monopole charge in self-interacting Kalb-Ramond field 
%: shadow and gravitational lensing
}

\author{Mohsen Fathi
\orcidlink{0000-0002-1602-0722}}
\email{mohsen.fathi@ucentral.cl}
\affiliation{Centro de Investigaci\'{o}n en Ciencias del Espacio y F\'{i}sica Te\'{o}rica, Universidad Central de Chile, La Serena 1710164, Chile}
%\affiliation{Grupo de Ciencias del Espacio y F\'{i}sicas,\\
%Universidad Central de Chile, Toesca 1783, Santiago 8320000, Chile}
%\affiliation{Facultad de Ingeniería y Arquitectura, Universidad Central de Chile,\\Av. Santa Isabel 1186, Santiago 8330563, Chile}

\author{Ali \"Ovg\"un
\orcidlink{0000-0002-9889-342X}
}
\email{ali.ovgun@emu.edu.tr}

\affiliation{Physics Department, Eastern Mediterranean
University, Famagusta, 99628 North Cyprus, via Mersin 10, Turkiye}

%%%%abstract
\begin{abstract}
In this study, we explore a static, spherically symmetric black hole solution in the context of a self-interacting Kalb-Ramond field coupled with a global monopole. By incorporating the effects of Lorentz-violating term $\ell$ and the monopole charge $\eta$ in the KR field, we derive the modified gravitational field equations and analyze the resulting black hole spacetime. The obtained solution exhibits deviations from the Schwarzschild metric with topological defect, as it is influenced by the monopole charge and self-interaction potential. We investigate the thermodynamic properties of the black hole, including its Hawking temperature, entropy, and specific heat, revealing novel stability conditions. Additionally, we perform solar system tests such as perihelion precession, gravitational redshift, light deflection, and time delay of signals to impose constraints on the Lorentz-violating parameter and monopole charge. Our findings suggest that these parameters have to be significantly small, although there are different constraints imposed by individual tests, ranging from $10^{-9}\leq|\ell|\leq 10^{-4}$ and $10^{-9}\leq\eta\leq 10^{-6}\, \mathrm{m}^{-1}$.
\end{abstract}

\date{\today}



\maketitle
%\tableofcontents

%%%%%%%%%%%%%%%%%%%%%%%%%%%%%%%%%%Sect. I
\section{Introduction}

Lorentz symmetry plays a pivotal role in modern physics, ensuring the consistency of the fundamental laws of nature across all inertial reference frames. This symmetry has been validated through extensive experimental observations. However, various theoretical approaches propose that Lorentz symmetry could be violated at high energy scales. Prominent examples of such theories include string theory \cite{Kostelecky1989a}, loop quantum gravity \cite{Alfaro2002}, Ho\v{r}ava-Lifshitz gravity \cite{Horava2009a}, non-commutative field theory \cite{Carroll2001}, as well as others such as Einstein-\ae ther theory \cite{Jacobson2001}, massive gravity \cite{Dubovsky2005}, $f(T)$ gravity \cite{Bengochea2009}, and very special relativity \cite{Cohen2006}. Understanding the breakdown of Lorentz symmetry is crucial for exploring the high-energy characteristics of physical systems and the nature of gravity \cite{Lehnert:2006mn,Kostelecky:2021bsb,Lehnert:2004be,Diaz:2013saa,Cambiaso:2012vb,Seifert:2010uu}.

Lorentz symmetry breaking (LSB) can occur in two distinct forms: explicit and spontaneous. Explicit LSB arises when the Lagrangian does not respect Lorentz invariance, meaning that the physical laws take different forms in different reference frames. On the other hand, spontaneous LSB occurs when the Lagrangian itself remains Lorentz invariant, but the vacuum state of the system does not exhibit Lorentz symmetry \cite{Gullu:2020qzu,Ovgun:2018xys,Oliveira:2018oha,Lambiase:2024uzy,Carleo:2022qlv,Lambiase:2017adh,AraujoFilho:2024ykw,Filho:2022yrk,Araujo:2024tiy,Nascimento:2023auz,Filho:2022yrk,Heidari:2024bvd,Hosseinifar:2024wwe,AraujoFilho:2024ctw,Maluf:2018jwc}. The Standard-Model extension (SME) \cite{Kostelecky2004a} offers a comprehensive framework for studying spontaneous LSB, with bumblebee models \cite{Kostelecky1989a,Kostelecky1989,Kostelecky1989b,Bailey2006,Bluhm2008a} being some of the simplest examples within this framework. These models involve a vector field, known as the bumblebee field, which acquires a nonzero vacuum expectation value (VEV), thus breaking the local Lorentz symmetry by selecting a specific direction in spacetime.

Several solutions have been studied in the context of bumblebee gravity. Casana et al.~\cite{Casana2018} discovered a static, spherically symmetric solution that has been extensively analyzed in relation to gravitational lensing \cite{Ovgun2018,Lambiase:2023zeo,Kuang:2022xjp,Lambiase:2023zeo,Mangut:2023oxa,Pantig:2024ixc}, quark stars \cite{Panotopoulos:2024jtn},  Hawking radiation \cite{Sakalli:2023pgn} and quasinormal modes \cite{Oliveira2021}. Maluf et al.~\cite{Maluf2021} extended the work by deriving an Anti-de Sitter-Schwarzschild-like solution. Xu et al.~\cite{Xu2023} identified new classes of static spherical bumblebee black holes. Investigations into rotating bumblebee black holes \cite{Ding2020a,Ding2021a} and their associated phenomena, such as gravitational wave propagation \cite{Liang2022,Amarilo2023}, have provided crucial insights into the effects of spontaneous LSB on gravitational behavior.

In addition to vector fields, rank-two antisymmetric tensor fields, such as the Kalb-Ramond (KR) field, have been explored as potential sources of LSB \cite{altschul_lorentz_2010}. The KR field, which arises in the context of bosonic string theory \cite{Kalb1974}, has been examined in various areas, including black hole physics \cite{Kao1996,Kar2003,Chakraborty2017,Junior:2024vdk,Zahid:2024ohn,Jumaniyozov:2024eah,Ditta:2024lnb,al-Badawi:2024pdx,Ortiqboev:2024mtk,Junior:2024ety,Ali:2023amn,Al-Badawi:2023xig,Rahaman:2023swt,Baruah:2023rhd}, cosmology \cite{Nair2022}, and braneworld scenarios \cite{Fu2012,Chakraborty2016}. When the KR field nonminimally couples to gravity and acquires a nonzero VEV, Lorentz symmetry is spontaneously broken. Lessa et al.~\cite{Lessa2020} provided a static, spherically symmetric solution within this framework, which was later explored for the motion of massive and massless particles \cite{Atamurotov2022}. The gravitational deflection of light and the shadow of rotating KR black holes were further studied by Kumar et al.~\cite{Kumar2020c}. Solutions for traversable wormholes \cite{Lessa2021,Maluf2022} and their implications for Bianchi type I cosmology \cite{Maluf2022a} have also been investigated. 
Recently, new exact solutions have been found for static and spherically symmetric spacetimes, both with and without the cosmological constant, in the presence of a nonzero VEV of the KR field, as reported in Refs. \cite{yang_static_2023,liu_static_2024,duan_electrically_2024}.

In this study, we build upon the aforementioned works by incorporating a nonzero VEV of the KR field and examining its interaction with a global monopole charge. Global monopoles, which arise as topological defects due to the spontaneous breaking of global symmetries, particularly in the context of early universe phase transitions, have been extensively explored in cosmology and black hole physics~\cite{Vilenkin1981,Vilenkin1982,barriola_gravitational_1989,Dadhich:1997mh}. These monopoles, resulting from an $\mathrm{O}(3) \to \mathrm{U}(1)$ symmetry breaking, are typically associated with spacetime singularities. However, when gravitational effects are taken into account, the properties of global monopoles are modified. The introduction of self-gravity resolves the self-energy divergence of monopoles, enhancing their physical relevance, especially in astrophysical contexts~\cite{Bronnikov}. Additionally, global monopoles formed during early universe phase transitions have been proposed as potential contributors to cosmic inflation~\cite{Preskill, Barriola}.

In the present work, we investigate the combined effects of spontaneous LSB in the KR field and the presence of global monopoles. We present a non-asymptotically flat solution that incorporates both phenomena, focusing on their implications for black hole thermodynamics and astrophysical observations. The interaction between these two phenomena provides valuable insights into the behavior of light and particles near black holes, offering potential observational consequences that can be tested in astrophysical settings.

This paper is organized as follows: In Sect.~\ref{sec:solution}, we present the field equations for the self-interacting KR field, followed by assigning the Lagrangian density of the global monopole as the matter Lagrangian. We then solve the field equations within a static, spherically symmetric metric ansatz to obtain the exterior solution of a static black hole in the KR background. Additionally, we provide a detailed analysis of the spacetime structure and its sensitivity to the spacetime parameters. In Sect.~\ref{sec:thermo}, we explore various thermodynamical aspects of the spacetime for the specific case where the black hole possesses only one horizon and derive corrections to the relations characteristic of the Schwarzschild black hole (SBH). In Sect.~\ref{sec:solarTests}, we perform the four standard Solar System tests on our newly obtained black hole spacetime. Within this context, we provide constraints on the orders of magnitude for the black hole parameters related to the Lorentz-violating and monopole charge parameters. We conclude in Sect. \ref{sec:conclusions}.

Throughout this study, we adopt the natural system of units by setting $\kappa = 8 \pi G = c = \hbar = 1$, and we work with the sign convention $(-,+,+,+)$. Furthermore, primes denote differentiation with respect to the argument of the corresponding functions wherever they appear.

%\textcolor{Orange}{\textit{\underline{abbreviations}}: KR (Kalb-Ramond), SBH (Schwarzschild black hole), EBH (extremal black hole), Lorentz symmetry breaking (LSB), Standard-Model extension (SME),vacuum expectation value (VEV)}

%%%%%%%%%%%%%%%%%%%%%%%%%%%%%%%%%Sect. II
\section{Neutral black hole solution in KR theory with global monopole constituent}\label{sec:solution}

The Einstein-Hilbert action for a gravity theory minimally coupled to the KR field is expressed as \cite{altschul_lorentz_2010,lessa_modified_2020}
\begin{equation}
\mathcal{S} = \frac{1}{2}\int d^4 x\sqrt{-g}\Biggl[
R - \frac{1}{6} H^{\mu\nu\rho} H_{\mu\nu\rho} - V\left(B^{\mu\nu} B_{\mu\nu}\right) + \xi_2 B^{\rho\mu} {B^\nu}_{\mu} R_{\rho\nu} + \xi_3 B^{\mu\nu} B_{\mu\nu} R
\Biggr] + \int d^4 x\sqrt{-g}\,\mathcal{L}_\mathrm{M},
\label{eq:action}
\end{equation}
where $\xi_2$ and $\xi_3$ are coupling constants that describe the interaction between gravity and the KR field. The term $H_{\mu\nu\rho} \equiv \partial_{[\mu} B_{\nu\rho]}$ represents the strength of the KR field, with $B_{\mu\nu} = \partial_\mu B_\nu - \partial_\nu B_\mu$ being the field strength tensor associated with the field $B^\mu$. The self-interaction potential $V(X)$, with $X = B^{\mu\nu} B_{\mu\nu}$, ensures the theory's invariance under local Lorentz transformations. Additionally, $\mathcal{L}_\mathrm{M}$ denotes the matter field Lagrangian density.

By varying the action in Eq.~\eqref{eq:action} with respect to the metric $g^{\mu\nu}$, the gravitational field equations are derived as 
\begin{equation}
R_{\mu\nu} - \frac{1}{2} g_{\mu\nu} R = T^{\mathrm{M}}_{\mu\nu} + T^{\mathrm{KR}}_{\mu\nu} = T^{\mathrm{tot}}_{\mu\nu},
\label{eq:field_0}
\end{equation}
where $T^{\mathrm{tot}}_{\mu\nu}$ is the total energy-momentum tensor, comprising the matter field energy-momentum tensor $T^{\mathrm{M}}_{\mu\nu}$ and the KR field energy-momentum tensor $T^{\mathrm{KR}}_{\mu\nu}$. The KR field energy-momentum tensor is given by
\begin{eqnarray}
T_{\mu\nu}^{\mathrm{KR}} &=& \frac{1}{2} H_{\mu\alpha\beta} {H_{\nu}}^{\alpha\beta} - \frac{1}{12} g_{\mu\nu} H^{\alpha\beta\rho} H_{\alpha\beta\rho} + 2 V'(X) B_{\alpha\mu} {B^\alpha}_\nu - g_{\mu\nu} V(X) \nonumber \\
&& + \xi_2 \Bigg[ \frac{1}{2} g_{\mu\nu} B^{\alpha\gamma} {B^{\beta}}_\gamma R_{\alpha\beta} - {B^\alpha}_\mu {B^{\beta}}_\nu R_{\alpha\beta} - B^{\alpha\beta} B_{\nu\beta} R_{\mu\alpha} - B^{\alpha\beta} B_{\mu\beta} R_{\nu\alpha} \nonumber \\
&&  + \frac{1}{2} \nabla_\alpha \nabla_\mu (B^{\alpha\beta} B_{\nu\beta}) + \frac{1}{2} \nabla_\alpha \nabla_\nu (B^{\alpha\beta} B_{\mu\beta}) - \frac{1}{2} \Box ({B_\mu}^\gamma B_{\nu\gamma})  - \frac{1}{2} g_{\mu\nu} \nabla_\alpha \nabla_\beta (B^{\alpha\gamma} {B^{\beta}}_\gamma) \Bigg],
\label{eq:TKR}
\end{eqnarray}
where $\Box \equiv \nabla^\alpha \nabla_\alpha$. Conservation of the total energy-momentum tensor, $T^{\mathrm{tot}}_{\mu\nu}$, can be verified using the Bianchi identities.

To allow for a non-vanishing VEV of the KR field, i.e., $\langle B_{\mu\nu} \rangle = b_{\mu\nu}$, we assume the potential takes the form $V = V(B^{\mu\nu} B_{\mu\nu} \pm b^2)$, where the $\pm$ sign ensures the positiveness of $b^2$ \cite{bluhm_spontaneous_2008,altschul_lorentz_2010,lessa_modified_2020}. The VEV configuration is determined by the constant norm condition $b^{\mu\nu} b_{\mu\nu} = \mp b^2$. Under this configuration, the field equations can be rewritten as \cite{yang_static_2023,liu_static_2024,duan_electrically_2024}
\begin{eqnarray}
R_{\mu\nu} &=& T_{\mu\nu}^{\mathrm{M}} - \frac{1}{2} g_{\mu\nu} T^{\mathrm{M}} + V'(Y) + \xi_2 \Bigg[ g_{\mu\nu} b^{\alpha\gamma} {b^\beta}_\gamma R_{\alpha\beta} - {b^\alpha}_\mu {b^\beta}_\nu R_{\alpha\beta} \nonumber \\
&& - b^{\alpha\beta} b_{\mu\beta} R_{\nu\alpha} - b^{\alpha\beta} b_{\nu\beta} R_{\mu\alpha} + \frac{1}{2} \nabla_\alpha \nabla_\mu (b^{\alpha\beta} b_{\nu\beta}) + \frac{1}{2} \nabla_\alpha \nabla_\nu (b^{\alpha\beta} b_{\mu\beta}) 
 - \frac{1}{2} \Box ({b_\mu}^\gamma b_{\nu\gamma}) \Bigg],
\label{eq:field_1}
\end{eqnarray}
where $T^{\mathrm{M}} = g^{\mu\nu} T^{\mathrm{M}}_{\mu\nu}$ and $Y = 2 b_{\mu\alpha} {b_\nu}^\alpha + b^2 g_{\mu\nu}$.

In this work, the matter Lagrangian density $\mathcal{L}_\mathrm{M}$ in Eq.~\eqref{eq:action} corresponds to a global monopole, defined as %\cite{barriola_gravitational_1989}
\begin{equation}
\mathcal{L}_{\mathrm{M}} \equiv \mathcal{L}_{(\mathrm{GM})} = \frac{1}{2} \partial_\mu \varphi^a \partial^\mu \varphi^a - \frac{\lambda}{4} \left( \varphi^a \varphi^a - \eta^2 \right)^2,
\label{eq:LGM}
\end{equation}
where $\varphi^a$ is a scalar triplet field ($a = 1, 2, 3$) representing the monopole configuration. The model exhibits global $\mathrm{O}(3)$ symmetry, spontaneously broken to $\mathrm{U}(1)$. Here, $\lambda$ is the self-coupling constant, and $\eta$ is the symmetry-breaking scale (monopole charge) with dimensions $[\eta] = \mathrm{length}^{-1}$. The corresponding energy-momentum tensor is given by
\begin{equation}
T^{\mathrm{M}}_{\mu\nu} \equiv T_{\mu\nu}^{\text{(GM)}} = \partial_\mu \varphi^a \partial_\nu \varphi^a - g_{\mu\nu} \left[ \frac{1}{2} \partial^\rho \varphi^a \partial_\rho \varphi^a - \frac{\lambda}{4} \left( \varphi^a \varphi^a - \eta^2 \right)^2 \right].
\label{eq:TGM}
\end{equation}
Focusing on a static, spherically symmetric spacetime, we adopt the metric ansatz
\begin{equation}
ds^2 = -A(r) dt^2 + B(r) dr^2 + r^2 d\theta^2 + r^2 \sin^2\theta \, d\phi^2,
\label{eq:metric_0}
\end{equation}
in Schwarzschild coordinates $(t, r, \theta, \phi)$. Assuming a spherically symmetric global monopole configuration, the scalar field is parametrized as $\varphi^a = \eta f(r) x^a / r$, with $x^a x^a = r^2$ and $f(r)$ being an arbitrary radial function. In this spacetime, the components of the energy-momentum tensor in Eq.~\eqref{eq:TGM} are obtained as 
\begin{subequations}
\begin{align}
T_{tt}^{(\mathrm{GM})} &= -\frac{1}{4} \eta^2 A(r) \left[\frac{2 f'(r)^2}{B(r)} + \frac{4 f(r)^2}{r^2} + \eta^2 \lambda \left(f(r)^2 - 1\right)^2 \right], \\
T_{rr}^{(\mathrm{GM})} &= \frac{1}{4} \eta^2 B(r) \left[\frac{2 f'(r)^2}{B(r)} + \frac{4 f(r)^2}{r^2} + \eta^2 \lambda \left(f(r)^2 - 1\right)^2 \right], \\
T_{\theta\theta}^{(\mathrm{GM})} &= \csc^2\theta \, T_{\phi\phi}^{(\mathrm{GM})} = \frac{1}{4} \eta^2 r^2 \left[\frac{2 f'(r)^2}{B(r)} + \eta^2 \lambda \left(f(r)^2 - 1\right)^2 \right].
\label{eq:TGM_1}
\end{align} 
\end{subequations}
For the specific case $ f(r) = 1 $, the above tensor simplifies to $ T_{\mu\nu}^{(\mathrm{GM})} = \mathrm{diag}\left(-{\eta^2 A(r)}/{r^2}, {\eta^2 B(r)}/{r^2}, 0, 0\right) $, and $ T^{(\mathrm{GM})} = {2 \eta^2}/{r^2} $. Using this, we find $ T_{\mu\nu}^{(\mathrm{GM})} - \frac{1}{2} g_{\mu\nu} T^{(\mathrm{GM})} = (0, 0, -\eta^2, -\eta^2 \sin^2\theta) $. With the help of Eqs.~\eqref{eq:field_1} and the spacetime metric \eqref{eq:metric_0}, we derive the following field equations \cite{Gullu:2020qzu}
\begin{eqnarray}
&&\frac{2A''(r)}{A(r)} - \frac{A'(r)}{A(r)}\frac{B'(r)}{B(r)} - \frac{A'(r)^2}{A(r)^2} + \frac{4}{r}\frac{A'(r)}{A(r)} = 0, \label{eq:fieldEq_tt}\\
&& \frac{2A''(r)}{A(r)} - \frac{A'(r)}{A(r)}\frac{B'(r)}{B(r)} - \frac{A'(r)^2}{A(r)^2} - \frac{4}{r}\frac{B'(r)}{B(r)} = 0, \label{eq:fieldEq_rr}\\
&& \frac{2A''(r)}{A(r)} - \frac{A'(r)}{A(r)}\frac{B'(r)}{B(r)} - \frac{A'(r)^2}{A(r)^2} + \frac{1+\ell}{\ell r} \left[\frac{A'(r)}{A(r)} - \frac{B'(r)}{B(r)} \right] 
- \Bigl[1 - b^2 r^2 V'(Y) \Bigr]\frac{2B(r)}{\ell r^2} + \frac{2(1-\ell)}{\ell r^2} + \eta^2 = 0, \label{eq:fieldEq_thetatheta}
\end{eqnarray}
where $ \ell \equiv \xi_2 b^2 / 2 $ is dimensionless. Following Ref.~\cite{yang_static_2023}, we assume $ V'(Y) = 0 $, which corresponds to the VEV being at the local minimum of the potential. For instance, this is achieved when $ V(X) = \frac{1}{2} \alpha X^2 $ with $ X = B^{\mu\nu}B_{\mu\nu} + b^2 $, where $ \alpha $ is a real coupling constant \cite{bluhm_spontaneous_2008}. 

Under these conditions, subtracting Eq.~\eqref{eq:fieldEq_rr} from Eq.~\eqref{eq:fieldEq_tt} yields $ A'(r)/A(r) = -B'(r)/B(r) $, leading to $ A(r) = B(r)^{-1} $. Subtracting Eq.~\eqref{eq:fieldEq_thetatheta} from Eq.~\eqref{eq:fieldEq_tt} provides the solution
\begin{equation}
A(r) = \frac{e^{-\frac{\ell \eta^2 r^2}{4(1-\ell)}}}{\eta(1-\ell)r} \left[\eta(1-\ell) c_1 + \sqrt{\pi\left(\frac{1}{\ell} - 1\right)} \, \mathrm{erf}_\ii\left(\frac{\eta r}{2\sqrt{\frac{1}{\ell} - 1}} \right) \right],
\label{eq:A(r)_0}
\end{equation}
where $ \mathrm{erf}_\ii(z) \equiv \mathrm{erf}(\ii z)/\ii $ is the imaginary error function, and $ c_1 $ is an integration constant. It has been shown that $ c_1 / 2 = M $ \cite{yang_static_2023}. Expanding the solution up to the second order of the monopole charge, we obtain
\begin{equation}
A(r) \approx \frac{1}{1-\ell} - \frac{2M}{r} + \frac{\ell M \eta^2 r}{2(1-\ell)} - \frac{\ell \eta^2 r^2}{6(1-\ell)^2} + \mathcal{O}(\eta^4).
\label{eq:A(r)_1}
\end{equation}
The solution is independent of the sign of $ \eta $, while the parameter $ \ell $ significantly impacts it. For $ \eta = 0 $, the solution reduces to
\begin{equation}
A(r) \approx \frac{1}{1-\ell} - \frac{2M}{r},
\label{eq:A(r)_2}
\end{equation}
which corresponds to a static uncharged black hole of mass $ M $ in a self-interacting KR field \cite{yang_static_2023}. This further simplifies to the Schwarzschild metric when the Lorentz-violating term vanishes. Due to experimental constraints on Lorentz-violating effects, $ \ell $ is expected to be small.

The Kretschmann scalar is given by:
\begin{multline}
R^{\alpha\beta\gamma\delta}R_{\alpha\beta\gamma\delta} = 
\frac{2 \eta^4 \ell^2}{3 (\ell-1)^4} + \frac{2 \eta^4 \ell^2 M}{(\ell-1)^3 r} + \frac{2 \eta^2 \ell^2 \bigl[3 \eta^2 (\ell-1) M^2 + 2\bigr]}{3 (\ell-1)^3 r^2} + \frac{4 \eta^2 \ell^2 M}{(\ell-1)^2 r^3} \\
+ \frac{4 \ell^2}{(\ell-1)^2 r^4} + \frac{16 \ell M}{(\ell-1) r^5} + \frac{48 M^2}{r^6},
\label{eq:KrtScl}
\end{multline}
which diverges at $ r = 0 $, indicating an intrinsic and non-removable singularity. The terms $ \sim r $ and $ \sim r^2 $ in $ A(r) $ resemble solutions with quintessential and cosmological terms (e.g., Ref.~\cite{fathi_study_2022}), though these arise from KR field contributions rather than cosmological effects. The general lapse function can be approximated as
\begin{equation}
A(r) \approx 1 - \frac{2M}{r} + l + \gamma r - k r^2,
\end{equation}
mimicking the Mannheim-Kazanas solution in conformal Weyl gravity \cite{mannheim_exact_1989}. However, these terms reflect spacetime curvature modifications due to the KR field and monopole charge, rather than cosmological contributions. To explore this further, we study the black hole's causal structure.

\begin{figure}[t]
    \centering
    \includegraphics[width=7cm]{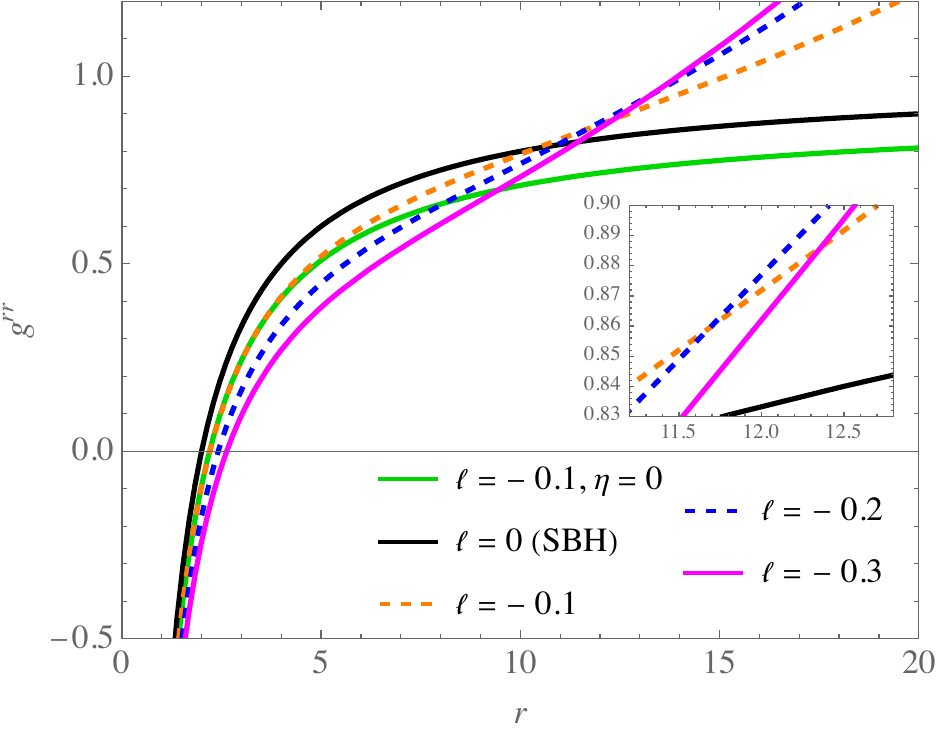} (a)\qquad
    \includegraphics[width=7cm]{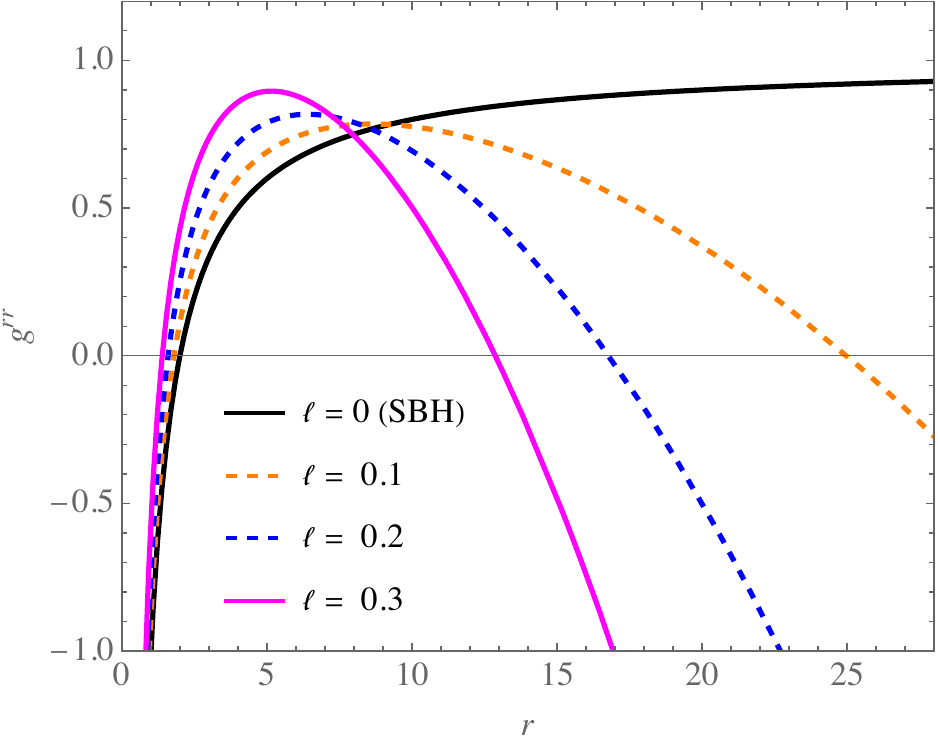} (b)
    \caption{Radial profile of $g^{rr}=A(r)$ for $\eta=0.3$, with (a) $\ell\leq0$ and an additional curve for $\eta=0$ (SBH with a topological defect), and (b) $\ell\geq0$. Unless otherwise noted, axes lengths are measured in units of the black hole mass $M$.}
    \label{fig:A(r)}
\end{figure}

\begin{figure}[h]
    \centering
    \includegraphics[width=8cm]{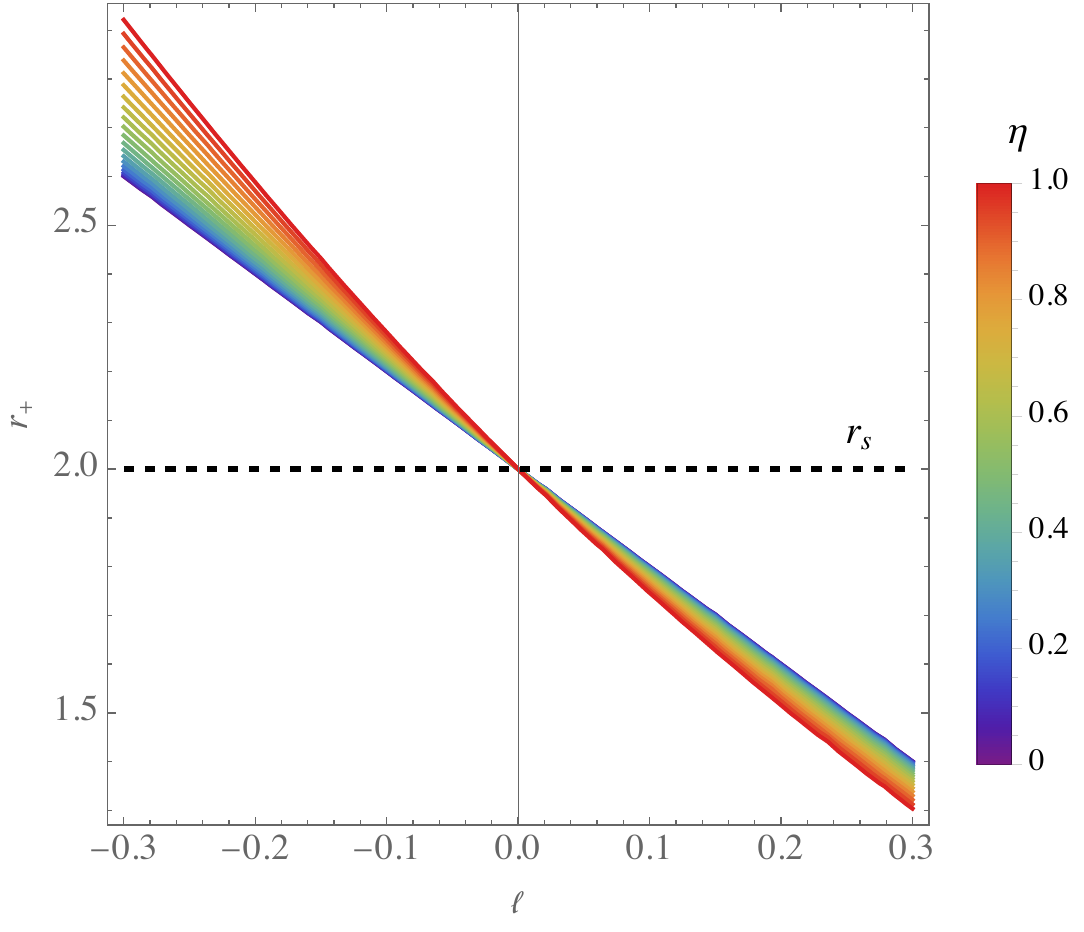} 
    \caption{Profiles of $r_+$ versus $\ell$ for $0\leq\eta\leq1$.}
    \label{fig:A(r)=0}
\end{figure}

To examine the causal structure of the spacetime characterized by the lapse function \eqref{eq:A(r)_1}, the horizon locations can be identified by solving the equation $g^{rr}=A(r)=0$. This equation yields three solutions:  
\begin{eqnarray}
    r_1 &=& (1-\ell)M-\frac{24(1-\ell)^2}{\ell\eta^2}\sqrt{\frac{\zeta_2}{3}}\,\cos\left(\frac{1}{3}\arccos\left(\frac{3\zeta_3}{\zeta_2}\sqrt{\frac{3}{\zeta_2}}\,\right)\right),\label{eq:r1}\\
    r_2 &=& (1-\ell)M-\frac{24(1-\ell)^2}{\ell\eta^2}\sqrt{\frac{\zeta_2}{3}}\,\cos\left(\frac{1}{3}\arccos\left(\frac{3\zeta_3}{\zeta_2}\sqrt{\frac{3}{\zeta_2}}\,\right)-\frac{4\pi}{3}\right),\label{eq:r2}\\
    r_3 &=& (1-\ell)M-\frac{24(1-\ell)^2}{\ell\eta^2}\sqrt{\frac{\zeta_2}{3}}\,\cos\left(\frac{1}{3}\arccos\left(\frac{3\zeta_3}{\zeta_2}\sqrt{\frac{3}{\zeta_2}}\,\right)-\frac{2\pi}{3}\right),\label{eq:r3}
\end{eqnarray}
where 
\begin{subequations}
    \begin{align}
        & \zeta_2 = \frac{\ell \bigl[\eta ^2 (1-\ell) \ell M^2+2\bigr]\eta^2}{48 (1-\ell)^3},\\
        & \zeta_3 = -\frac{ \ell^2  \bigl[\eta ^2 (1-\ell) \ell M^2-3\bigr]M \eta^4}{1728 (1-\ell)^4}.
    \end{align}
    \label{eq:zeta23}
\end{subequations}
For $\ell \leq 0$, the conditions $r_1>0$ and $r_2=r_3^*\in\mathbb{C}$ hold, indicating that the black hole possesses a single horizon, $r_+=r_1$. Conversely, when $\ell>0$, we find $r_1<0$ and $0<r_3<r_2$. In this case, the black hole has an event horizon at $r_+=r_3$ and an outer horizon at $r_{++}=r_2$. Notably, since the theory lacks a cosmological constant, the existence of $r_{++}$ arises from the $\sim r^2$ terms in the lapse function. This behavior results from the interplay between the monopole charge, the KR field, and their interaction with spacetime, mimicking the effects of a cosmological constant. At the $r_+$ surface, infinite redshift occurs, whereas infinite blueshift is observed at $r_{++}$. Therefore, $r_{++}$ can be interpreted as an exterior Cauchy horizon, beyond which spacetime predictability fails. These aspects can be further illustrated by examining the radial profile of $g^{rr}=A(r)$, as shown in Fig. \ref{fig:A(r)}.  
The plots reveal that for $\ell<0$, the SBH represents the lower bound, and the black hole becomes larger than the SBH. For $\ell\geq0$, the SBH serves as the upper bound, with $r_+\leq r_s=2M$. Additionally, the spacetime asymptotically resembles Anti-de Sitter (AdS) spacetime for $\ell<0$ and de Sitter (dS) spacetime for $\ell>0$. The event horizon's dependence on $\ell$ and the monopole charge is shown in Fig. \ref{fig:A(r)=0}.  
The profiles demonstrate that as $\ell$ transitions from negative to positive values, the black hole size diminishes continuously, with the SBH being the limiting case for $\ell<0$ and $\ell>0$. Moreover, for $\ell<0$, increasing $\eta$ enlarges the black hole, whereas for $\ell>0$, the opposite trend is observed.  

\begin{figure}[h]
    \centering
    \includegraphics[width=7cm]{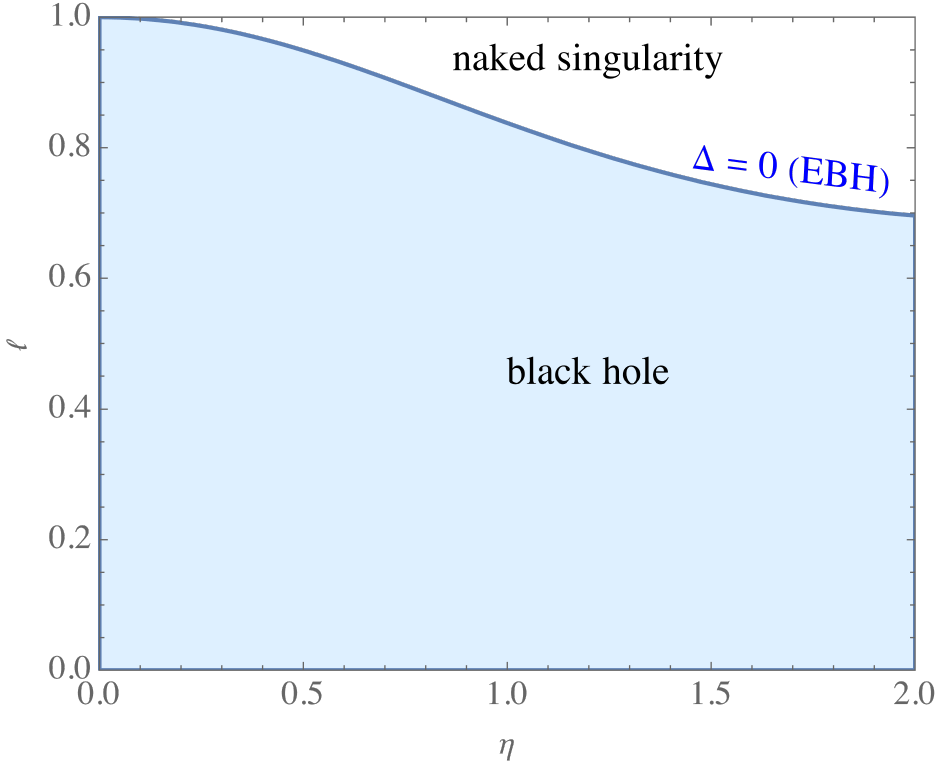} 
    \caption{Relationship between $\ell$ and $\eta$ for $\Delta>0$ (blue region), with the $\Delta=0$ curve (EBH) separating it from the naked singularity region ($\Delta<0$).}
    \label{fig:Delta=0}
\end{figure}

For $\ell\leq0$, a black hole exists for all values of the monopole charge. In contrast, for $\ell>0$, the horizons can merge into an extremal black hole (EBH) or vanish entirely, leaving a naked singularity. This occurs when the discriminant $\Delta=16(\zeta_2^3-27\zeta_3^2)$ of the cubic $A(r)=0$ for $\ell>0$ satisfies $\Delta=0$. When $\Delta>0$, the cubic equation has three real roots, with one negative and two positive. For $\Delta=0$, the positive roots degenerate, forming an EBH with a single horizon. If $\Delta<0$, the cubic equation yields one real negative root and two complex conjugates, eliminating the horizons and leaving a naked singularity. Figure \ref{fig:Delta=0} illustrates $\Delta$ as a function of $\eta$ and $\ell$ for $\ell>0$.  
The curve corresponding to $\Delta=0$ identifies the EBH. Solving $\Delta=0$ yields the critical value of $\eta$ for the EBH:  
\begin{equation}
\eta_*^2=\frac{3-4 \sqrt{3} M (1-\ell)^2-\sqrt{8 M (1-\ell)^2 \left[6 M (1-\ell)^2-7 \sqrt{3}\right]+9}}{2 M^2 (1-\ell) \ell },
    \label{eq:etastar}
\end{equation}
which denotes the value of $\eta$ at the EBH. However, as shown in Fig. \ref{fig:Delta=0}, this condition corresponds to larger $\ell$ values that lack physical relevance in the theory.

%%%%%%%%%%%%%%%%%Sect. III
\section{Thermodynamics of the black hole}\label{sec:thermo}

Black hole thermodynamics reveals that black holes follow the four thermodynamic laws \cite{bardeen_four_1973,kastor_enthalpy_2009} and exhibit phase structures akin to classical thermodynamic systems \cite{wei_insight_2015,kubiznak_black_2017,yang_kinetics_2022}, providing important insights into gravity and black hole physics. Asymptotically AdS-like black holes are particularly significant due to their connection to the AdS/CFT correspondence and their various phase transitions. In contrast, asymptotically dS-like black holes, which are characterized by two distinct temperatures, remain in non-equilibrium states, presenting unique challenges. Although some progress has been made in studying their thermodynamics \cite{mbarek_reverse_2019}, many aspects are still not fully understood. In this section, we focus on the fundamental thermodynamic properties of the asymptotically AdS-like black hole corresponding to the case of $\ell \leq 0$, for simplicity in our analysis.

\begin{figure}[h]
    \centering
    \includegraphics[width=8.3cm]{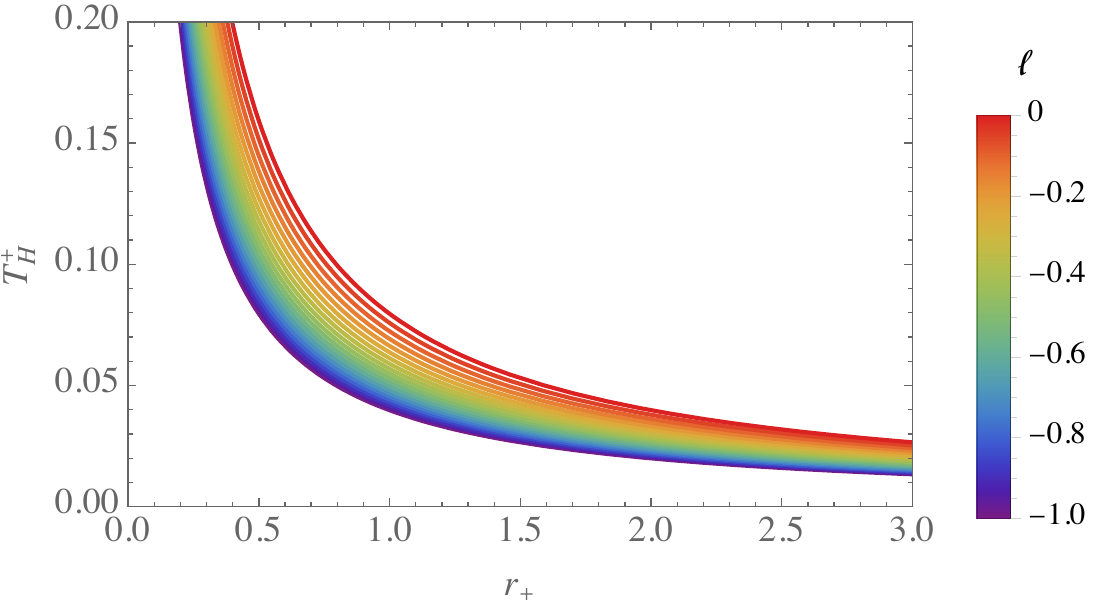} (a)
    \includegraphics[width=8.3cm]{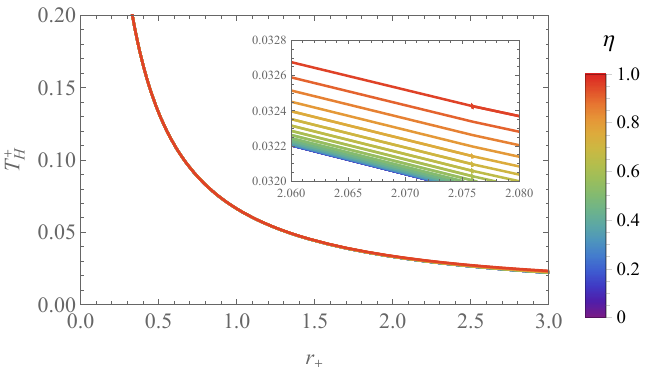} (b)
    \caption{The behavior of the Hawking temperature for the AdS-like black hole within $\ell \leq 0$, plotted for (a) $\eta = 0.3$, and (b) $\ell = -0.2$.}
    \label{fig:TH}
\end{figure}

To investigate this, we begin by solving the equation $A(r_+)=0$ for the lapse function \eqref{eq:A(r)_1} to determine the black hole mass, which gives
\begin{equation}
M = \frac{r_+ \Bigl[\ell \left(\eta ^2 r_+^2+6\right)-6\Bigr]}{3 (1-\ell) \Bigl[\ell \left(\eta ^2 r_+^2+4\right)-4\Bigr]}.
    \label{eq:Mr+}
\end{equation}
Since the black hole is static and spherically symmetric with an effective long-distance term $l_{\mathrm{eff}}=\frac{1}{2}\ell\eta^2/(1-\ell)^2$, we can associate the black hole mass $M$ with the enthalpy of the black hole. In this case, the thermodynamic pressure is given by $P=-\frac{1}{8\pi}l_{\mathrm{eff}}$, and thus, the enthalpy is related to the $\ell\eta^2$ coefficient, which accounts for both the Lorentz-violating parameter and the monopole charge. Since $\ell$ is dimensionless, we can express the enthalpy as a function of entropy $S$ and pressure $P$, in the form $M\equiv M(S,P)$. Consequently, the first law of thermodynamics for the AdS-like black hole is written as
\begin{equation}
dM = T_{\mathrm{H}}^+ dS+V dP,
    \label{eq:1st_0}
\end{equation}
where $T_{\mathrm{H}}^+$ is the Hawking temperature of the horizon and $V$ represents the thermodynamic volume. Using the line element \eqref{eq:metric_0} along with Eqs. \eqref{eq:A(r)_1} and \eqref{eq:Mr+}, the Hawking temperature can be computed as
\begin{equation}
T_{\mathrm{H}}^+ = \frac{A'(r_+)}{4\pi} = \frac{6 \eta ^2 (1-\ell) \ell r_+^2 - 24 (1-\ell)^2 - \eta ^4\ell^2 r_+^4}{24 \pi  (1-\ell)^2 r_+ \Bigl[\ell \left(\eta ^2 r_+^2+4\right)-4\Bigr]}.
    \label{eq:TH_0}
\end{equation}
For $\ell=0$, the SBH's Hawking temperature is recovered, which is given by $T_{\mathrm{H}}^+ = 1/(4\pi r_s)$. In Fig. \ref{fig:TH}, we plot the behavior of $T_{\mathrm{H}}^+$ as a function of $r_+$ for specific values of $\ell$ and $\eta$, considering $\ell\leq 0$.
The impact of the Lorentz-violating parameter on the Hawking temperature is evident from the diagrams. It can be inferred that for fixed $\eta$, increasing the $\ell$-parameter leads to an increase in temperature. Consequently, the SBH (i.e., $\ell = 0$) has the highest temperature. Similarly, for fixed $\ell$, an increase in the monopole charge results in a higher temperature. In general, all profiles show a significant decrease in temperature as the black hole size increases.

\begin{figure}[h]
    \centering
    \includegraphics[width=8.3cm]{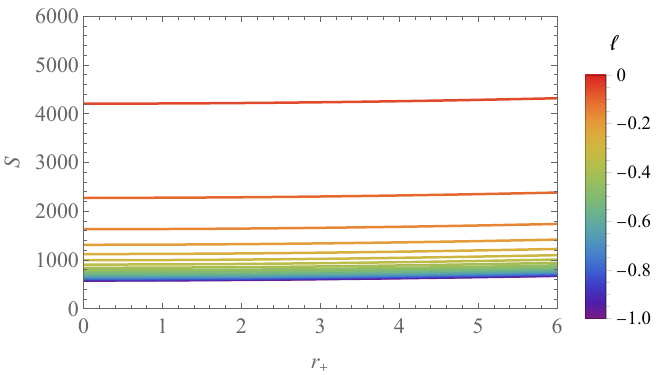} (a)
    \includegraphics[width=8.3cm]{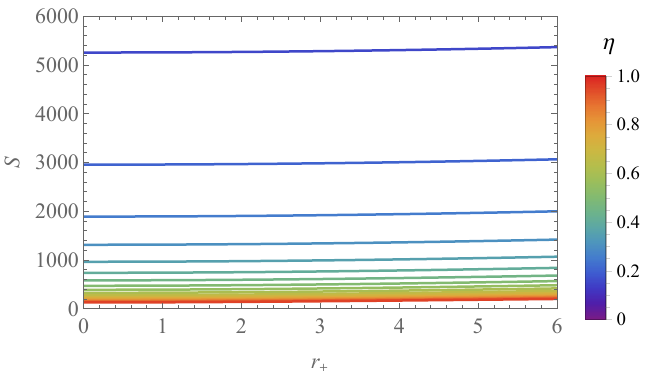} (b)
    \caption{Entropy behavior for the AdS-like black hole with $\ell \leq 0$ values, plotted for (a) $\eta = 0.3$, and (b) $\ell = -0.2$.}
    \label{fig:S}
\end{figure}

Using the first law of thermodynamics in Eq. \eqref{eq:1st_0} and the Hawking temperature in Eq. \eqref{eq:TH_0}, one can derive the black hole's entropy as 
\begin{eqnarray}
S &=& \int \left( \frac{dM}{T_{\mathrm{H}}^+} \right)_P = \int \frac{1}{T_{\mathrm{H}}^+} \left( \frac{\partial M}{\partial r_+} \right)_P dr_+ \nonumber \\
&=& - \frac{4 \pi (1-\ell) \ln \left( \eta^2 \ell r_+^2 + 4 - 4\ell \right)}{\eta^2 \ell} \nonumber \\
&\approx& \frac{\mathcal{A_+}}{4} - \frac{4 \pi (1-\ell) \ln (4 - 4\ell)}{\eta^2 \ell} + \mathcal{O}(r_+^3),
\label{eq:S_0}
\end{eqnarray}
where $\mathcal{A_+} = 4\pi r_+^2$ is the event horizon's area. This indicates that the Bekenstein-Hawking (B-H) area-entropy relation is not exactly satisfied, which is a consequence of the global monopole's presence. If we consider $\eta = 0$, the entropy reduces to $S = \mathcal{A_+}/4$, which aligns with the B-H formula. The behavior of entropy as a function of $r_+$ is depicted in Fig. \ref{fig:S} for variations in the $\ell$ and $\eta$ parameters.
As expected from the diagram, the SBH possesses the largest entropy. For fixed $\eta$, a decrease in the $\ell$-parameter leads to a decrease in entropy. In contrast, for a fixed $\ell$, a decrease in the $\eta$-parameter increases the entropy. The thermodynamic volume is calculated as 
\begin{eqnarray}
V &=& \left( \frac{\partial M}{\partial P} \right)_S = \left( \frac{\partial M}{\partial l_{\mathrm{eff}}} \right)_S \left( \frac{\partial l_{\mathrm{eff}}}{\partial P} \right)_S \nonumber \\
&=& \frac{16 \pi (1-\ell) r_+^3}{3 \left[ 4 - \ell \left( \eta^2 r_+^2 + 4 \right) \right]} \nonumber \\
&\approx& \frac{4 \pi}{3} r_+^3 + \frac{\ell \eta^2 \pi}{3} r_+^5 + \mathcal{O} \left( \ell^2, r_+^7 \right),
\label{eq:VT}
\end{eqnarray}
which reduces to the volume of a sphere of radius $r_+$ for $\eta \to 0$, as expected from the B-H criteria. Consequently, the Smarr formula is corrected as
\begin{equation}
M \approx 2 \left( T_{\mathrm{H}^+} S - V P \right) + \frac{4 \ln (2)}{\ell \eta^2 r_+} - \frac{2}{\eta^2 r_+} + \mathcal{O} \left( \ell^2 \right),
\label{eq:Smarr}
\end{equation}
for the black hole with monopole charge. It is evident that this result becomes singular for $\eta \to 0$. Therefore, to verify the Smarr formula for a neutral black hole in the KR field, we first set $\eta = 0$ in the integral of Eq. \eqref{eq:S_0}. This yields $S = \pi r_+^2$, $V ={4 \pi r_+^3}/{3}$, and $P = 0$, leading to $M = {r_+}/{(2 - 2\ell)}$, which corresponds to the mass of a neutral black hole in KR gravity \cite{yang_static_2023}, satisfying the Smarr formula.

\begin{figure}[h]
    \centering
    \includegraphics[width=8.3cm]{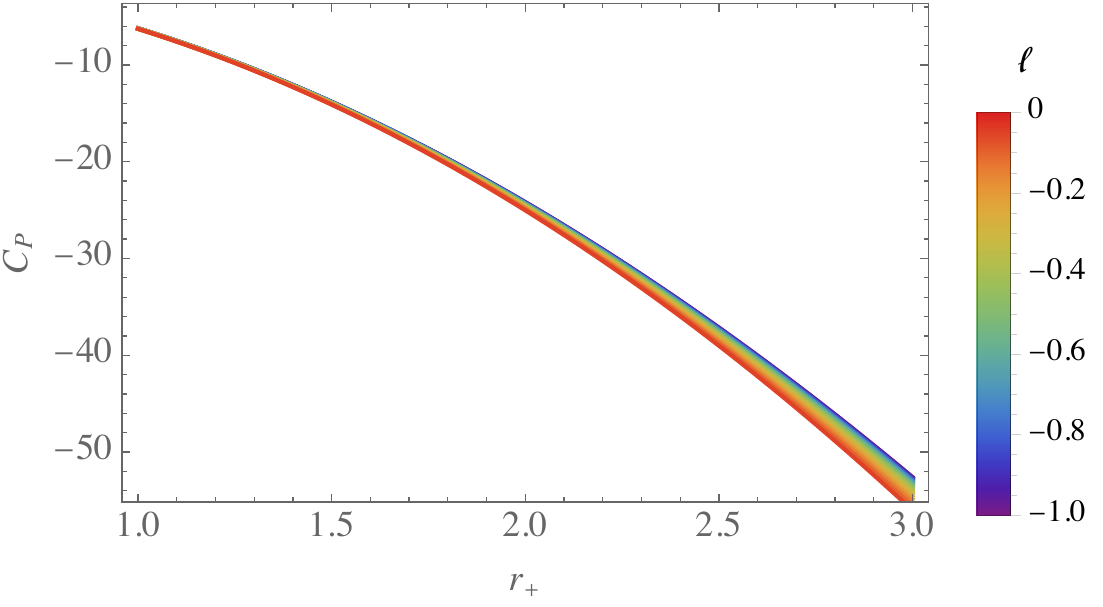} (a)
    \includegraphics[width=8.3cm]{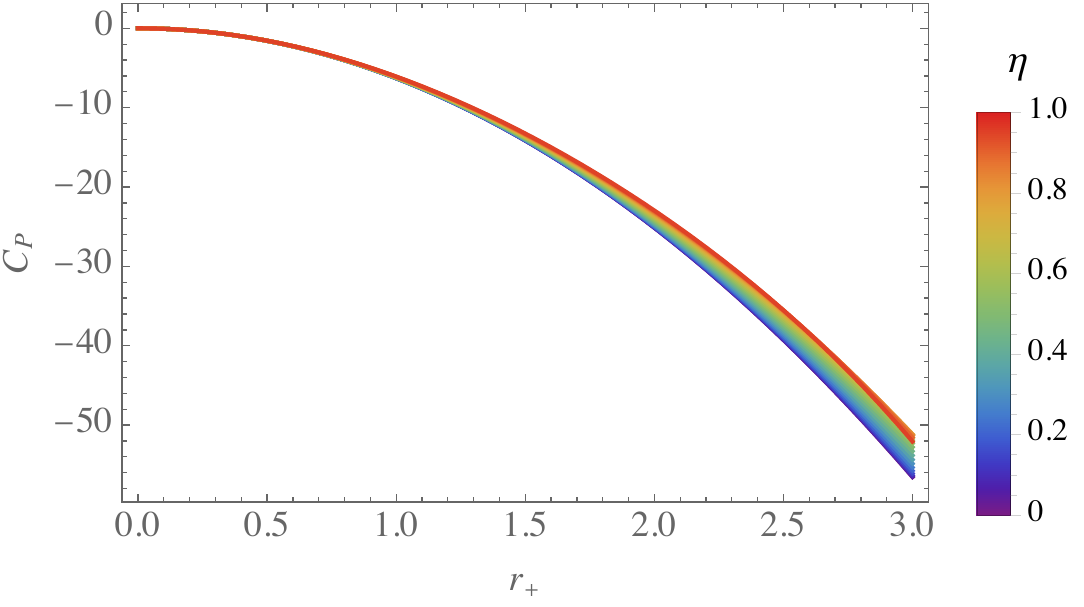} (b)
    \includegraphics[width=8.3cm]{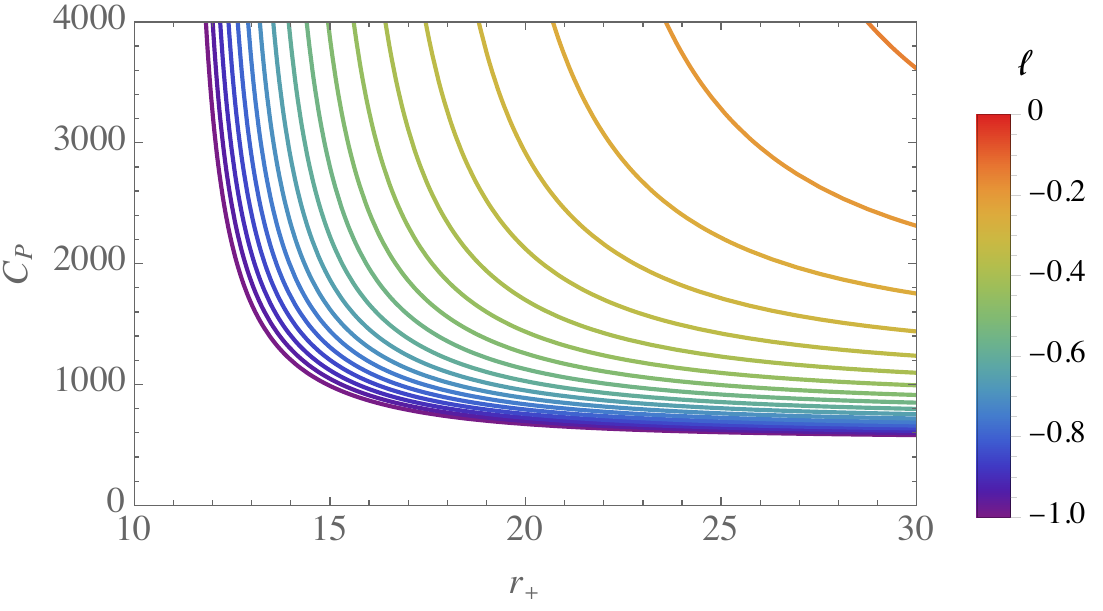} (c)
    \includegraphics[width=8.3cm]{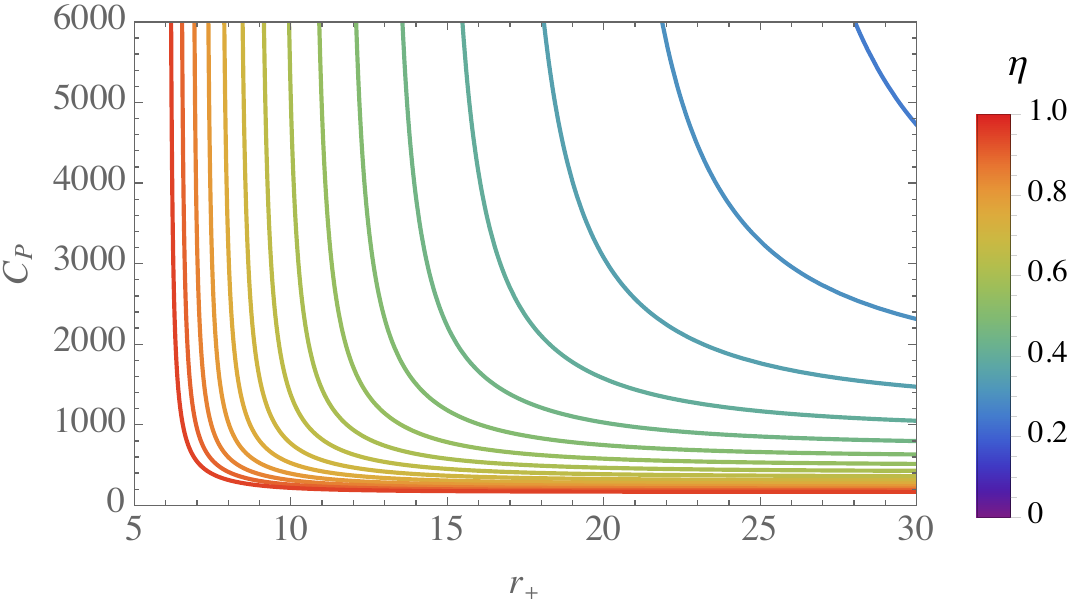} (d)
    \caption{The behavior of the specific heat within $ \ell \leq 0 $ values, plotted for (a,b) general relativistic black hole size range, and (c,d) larger black holes. The diagrams correspond to (a,c) $ \eta = 0.3 $, and (b,s) $ \ell = -0.2 $.}
    \label{fig:CP}
\end{figure}

Another thermodynamic quantity is the specific heat, which is important in analyzing the local stability of the black hole. This quantity is calculated as
\begin{eqnarray}
C_p &=& \left(\frac{\partial M}{\partial T_{\mathrm{H}^+}}\right)_P = \left(\frac{\partial M}{\partial r_+}\right)_P \left(\frac{\partial r_+}{\partial T_{\mathrm{H}^+}}\right)_P \nonumber \\
&=& \frac{8 \pi (1-\ell) r_+^2 \Bigl[6 \eta^2 (1-\ell) \ell r_+^2 - \eta^4 \ell^2 r_+^4 - 24 (1-\ell)^2\Bigr]}{\eta^6 \ell^3 r_+^6 - 6 \eta^4 (1-\ell) \ell^2 r_+^4 - 48 \eta^2 (1-\ell)^2 \ell r_+^2 + 96 (1-\ell)^3} \nonumber \\
&\approx& -2 \pi r_+^2 - \frac{1}{2} \pi \ell \eta^2 r_+^4 + \mathcal{O}(\ell^2),
\label{eq:Cp_0}
\end{eqnarray}
for the black hole with monopole charge, which results in $ C_p = -2 \pi r_+^2 $ in the limit of $ \eta \to 0 $, corresponding to the neutral black hole in the KR field \cite{yang_static_2023}. It is interesting to note that the contribution of the Lorentz-violating term only appears as an interaction with the monopole charge. Hence, for a neutral black hole, the $ \ell $-parameter does not contribute to the heat capacity, although it appears in both the mass and the Hawking temperature. In Fig. \ref{fig:CP}, the behavior of the heat capacity has been plotted versus changes in $ r_+ $, for different values of $ \ell $ and $ \eta $, within the domain $ \ell \leq 0 $.
As shown in the diagrams, the specific heat remains negative for black hole sizes within the range typically considered reasonable in general relativity, indicating local instability, akin to the behavior of the SBH. However, for sufficiently large black holes ($ r_+ \gtrsim 5M $), the specific heat becomes positive ($ C_p > 0 $), signifying stability. Notably, when $ C_p < 0 $, increasing the $ \ell $-parameter leads to a decrease in $ C_p $, while an increase in $ \eta $ causes $ C_p $ to rise. Conversely, when $ C_p > 0 $, an increase in $ \ell $ results in an increase in $ C_p $, while increasing $ \eta $ causes $ C_p $ to decrease.

\begin{figure}[h]
    \centering
    \includegraphics[width=8.3cm]{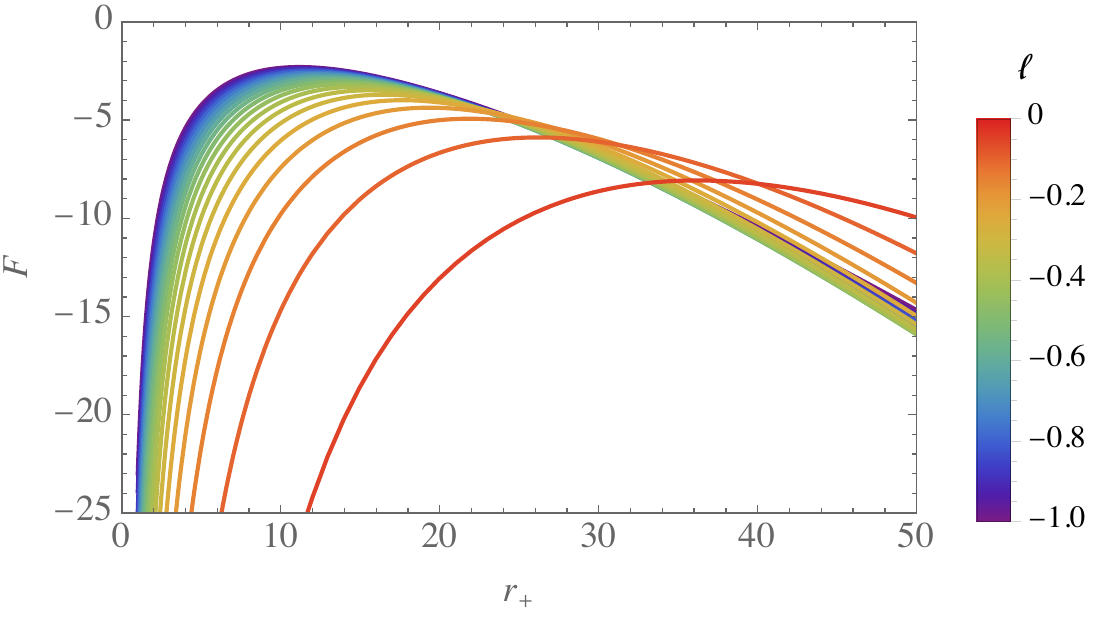} (a)
    \includegraphics[width=8.3cm]{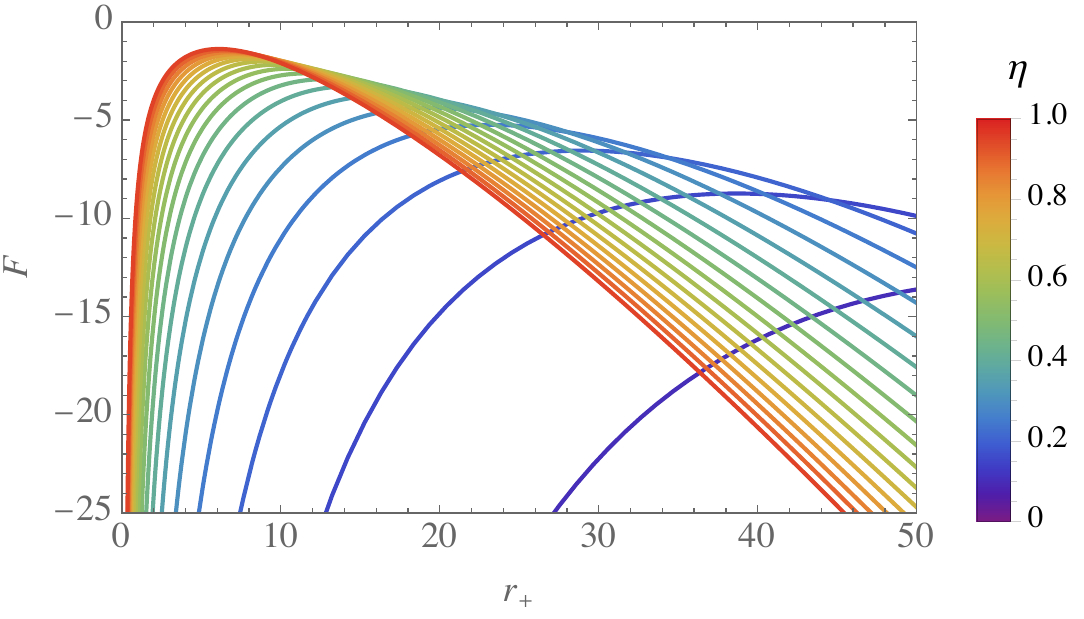} (b)
    \caption{The behavior of the Gibbs free energy within $ \ell \leq 0 $ values, plotted (a) $ \eta = 0.3 $, and (b) $ \ell = -0.2 $.}
    \label{fig:F}
\end{figure}

To analyze the global stability of black holes, one can consider the Gibbs free energy $ F $, in the sense that for $ F < 0 $, the black hole is globally stable, whereas for $ F > 0 $, it is globally unstable. The Gibbs free energy for the black hole under consideration is calculated as
\begin{eqnarray}
F &=& M - T_{\mathrm{H}}^+ S \nonumber \\
&=& \frac{\Bigl[24 (1-\ell)^2 + \eta^4 \ell^2 r_+^4 - 6 \eta^2 (1-\ell) \ell r_+^2\Bigr] \ln \left(\eta^2 \ell r_+^2 + 4 - 4\ell\right) + 2 \eta^2 \ell r_+^2 \Bigl[6 - \ell \left(\eta^2 r_+^2 + 6\right)\Bigr]}
{6 \eta^2 (1-\ell) \ell r_+ \Bigl[4 - \ell \left(\eta^2 r_+^2 + 4\right)\Bigr]} \nonumber \\
&\approx& \frac{r_+}{4(1-\ell)} + \frac{\ln (4 - 4 \ell)}{\eta^2 \ell r_+} + \mathcal{O}(\ell^2).
\label{eq:F_0}
\end{eqnarray}
Again, for $ S = \pi r_+^2 $, corresponding to $ \eta = 0 $, the above relation results in the Gibbs free energy for a neutral black hole in the KR field, which is $ F = {r_+}/{(4 - 4\ell)} $ \cite{yang_static_2023}. In Fig. \ref{fig:F}, the behavior of the Gibbs free energy has been plotted versus changes in the event horizon, for different values of $ \ell $ and $ \eta $.
As observed, the Gibbs free energy remains negative across the entire range of black hole sizes, suggesting that the black hole is globally stable. Additionally, we find that an increase in the $ \ell $-parameter leads to a decrease in the free energy, while an increase in the monopole charge results in an increase in the free energy. It is worth mentioning that the SBH is not only locally unstable but also globally unstable, as its free energy is always positive. However, despite its presence, the topological defect, which manifests itself by the Lorentz-violating parameter $ \ell $, cannot affect this instability for a Schwarzschild-like black hole in the KR field, as the free energy for this black hole remains positive.

%%%%%%%%%%%%%%%%%%%%Sect. IV
\section{Solar system tests}\label{sec:solarTests}

To analyze the observational implications of the obtained spacetime in the context of the solar system, it is essential to consider the weak-field approximation of the metric. In this regime, the radial coordinate $r$ is significantly larger than the characteristic scale of the black hole, making higher-order corrections in $r$ negligible. Therefore, we approximate the lapse function $A(r)$ as:
\begin{equation}
A(r) \approx 1 +\ell - \frac{2M}{r}  + \frac{\ell M \eta^2}{2}\, r + \mathcal{O}(\ell^2,r^2).
\label{eq:A(r)_app}
\end{equation}
This approximation is justified for the following reasons:
The term $-2M/r$ represents the dominant Schwarzschild potential, which governs gravitational interactions within the solar system.  The additional constant term $\ell$ introduces a small deviation due to Lorentz violation, which can be constrained by precise observations. The linear correction ${\ell M \eta^2 r}/{2}$ reflects the first-order effect of the monopole charge $\eta$, which remains significant in the weak-field regime. Higher-order terms, such as those proportional to  $r^2$, are negligible within the solar system due to the small values of $\ell$, and the relatively small scales involved compared to cosmological distances. In fact, given the smallness of the parameters, approximate approaches and perturbative methods can be applied to this black hole in the context of the solar system. Accordingly, using the above simplified form of the lapse function, we will now investigate the solar system tests, such as the perihelion precession, bending of light, and time delay of signals.

%%%%%%%%%%%%%%
\subsection{Perihelion precession}

\begin{figure}[h]
    \centering
    \includegraphics[width=10cm]{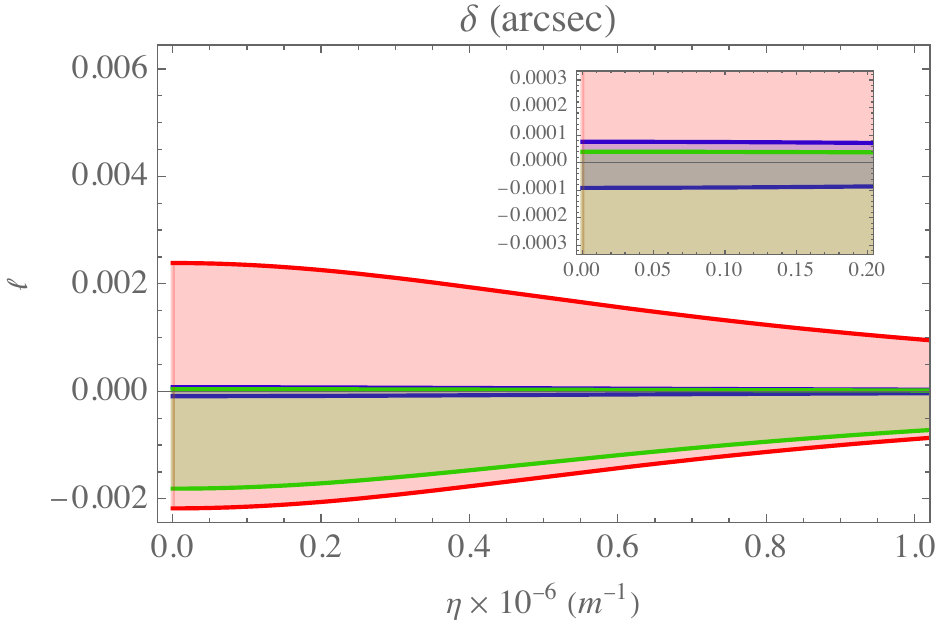} 
    \caption{Constraints on $\ell$ and $\eta$ based on perihelion precession data for Mercury (blue), Venus (green), and Earth (red) (see Ref.~\cite{cornbleet93} for data).}
    \label{fig:perihelionConst}
\end{figure}

An intuitive method to investigate this effect was introduced in Ref.~\cite{cornbleet93}, which funds our mathematical approach within this subsection. The core concept involves comparing Keplerian elliptic orbits in the Minkowski spacetime, described using Lorentzian coordinates, with those defined in Schwarzschild coordinates. This approach naturally highlights the general relativistic corrections. Consider the unperturbed Lorentzian metric 
\begin{equation}
\label{eq:lormet}
{d}s^2=-{ d}t^2+{ d}r^2
+r^2 { d}\theta^2+r^2 \sin^2\theta{ d}\phi^2,
\end{equation}
defined in the $(t,r,\theta,\phi)$ coordinates. We compare this with the metric \eqref{eq:metric_0}, now assumed to be expressed in the $(t',r',\theta,\phi)$ coordinates. The relationship between $(t,r)$ and $(t',r')$ is approximated as
\begin{subequations}
    \begin{align}
        & dt'=\left(1+\frac{\ell}{2}-\frac{M}{r}+\frac{1}{4} \eta ^2 \ell M r\right) dt,\\
        & dr'=\left(1-\frac{\ell}{2}+\frac{M}{r}-\frac{1}{4} \eta ^2 \ell M r\right) dr.
    \end{align}
    \label{eq:binomial}
\end{subequations}
In the invariant plane $\theta=\pi/2$, the area element in the Lorentzian framework is $d \mathcal{A}=\int_0^R r d r d\phi=\frac{1}{2}R^2 d\phi$, where $R$ represents the areal distance from the planet to the central mass. Thus, Kepler's second law can be written as
\begin{equation}
\label{eq:kep1}
\frac{{  d}\mathcal{A}}{{  d}t}=\frac{1}{2}R^2
\frac{{  d}\phi}{{  d}t}.
\end{equation}
In Schwarzschild coordinates, we have
\begin{eqnarray}
{ d}\mathcal{A}'&=&\int_0^R r{ d}r' { d}\phi= \int_0^R \left(r-\frac{\ell}{2} r+M-\frac{1}{4} \eta ^2 \ell M r^2\right){ d}r { d}\phi\\\label{eq:kep2}
&=&\frac{R^2}{2}\left(1-\frac{\ell}{2}+\frac{2 M}{R}-\frac{1}{6} \eta ^2 \ell M R\right){ d}\phi.
\end{eqnarray}
Using the transformations \eqref{eq:binomial}, Kepler's second law becomes
\begin{eqnarray}
\frac{{d}\mathcal{A}'}{{ d}t'}&=&\frac{1}{2}R^2
\left(1-\frac{\ell}{2}+\frac{2 M}{R}-\frac{1}{6} \eta ^2 \ell M R\right)
\frac{{ d}\phi}{{ d}t'}\\\nonumber
&=&\frac{1}{2}R^2
\left(1-\frac{\ell}{2}+\frac{2 M}{R}-\frac{1}{6} \eta ^2 \ell M R\right) 
\left(1-\frac{\ell}{2}+\frac{M}{R}-\frac{1}{4} \eta ^2 \ell M R\right)
\frac{{ d}\phi}{{ d}t}\\
&\simeq&\frac{1}{2}R^2
\left(1-\ell+\frac{3 M}{R}-\frac{2}{3} \eta ^2 \ell M^2\right)
\frac{{ d}\phi}{{ d}t}.
\label{eq:kep3}
\end{eqnarray}
Since this law must hold covariantly in all coordinate systems, Eqs.~\eqref{eq:kep1} and \eqref{eq:kep3} imply $d\phi'=(1-\ell+{3 M}/{R}-{2\eta ^2 \ell M^2}/{3})d\phi$. For an angular increment $\Delta\phi'$, we find
\begin{equation}
\int_0^{\Delta \phi'}{ d}\phi'= \int_0^{\Delta \phi=2\pi} \left(1-\ell+\frac{3 M}{R}-\frac{2}{3} \eta ^2 \ell M^2\right){ d}\phi,
\label{eq:incang}
\end{equation}
for one orbit. Given $R=l/(1+\varepsilon \cos \phi)$, where $\varepsilon$ is the eccentricity and $l$ the semi-latus rectum, this becomes
\begin{eqnarray}\nonumber
\Delta \phi'&=&2\pi \left(1-\ell-\frac{2}{3} \eta ^2 \ell M^2\right) +\frac{3M}{l}\int_0^{2\pi} \left(1+\varepsilon \cos \phi\right){d}\phi\\
&=& 2\pi +\Delta \phi_{\mathrm{gr}}+\Delta \phi_{\mathrm{KR}}+\Delta \phi_{\eta}, 
\end{eqnarray}
where
\begin{subequations}
\begin{align}
    & \Delta\phi_{\mathrm{gr}} = \frac{6\pi M}{l},\\
    & \Delta\phi_{\mathrm{KR}} = - 2\pi\ell,\\
    & \Delta\phi_{\eta} = -\frac{4\pi}{3} \ell \eta^2 M^2,
    \label{eq:Deltaphi}
    \end{align}
\end{subequations}
represent the general relativistic term and the corrections due to the KR field and the KR-monopole charge interaction. Using $M=M_{\odot}=1476.112$ m in our unit system, the perihelion advance in arcseconds per century is
\begin{equation} \label{eq:adpe}
\delta\equiv \Delta\phi'-2\pi= 5.73912\, \frac{\upsilon}{l}-1.296\,
\upsilon\ell - 1.88258\,   \upsilon \ell\eta^2,
\end{equation}
where $\upsilon$ is the number of orbits per year and $l$ is approximately $10^9$ m. Figure \ref{fig:perihelionConst} constrains the black hole parameters $\ell$ and $\eta$ based on perihelion advance observations.
From Fig.~\ref{fig:perihelionConst}, the optimal range for $\ell$ is $-0.0001 \leq \ell \leq 0.0001$ for $\eta\sim 10^{-6}\, \mathrm{m}^{-1}$. Additionally, since $\eta$ enters as a squared term, its sign does not affect the constraints, and for $\eta < 0$, a mirrored domain is obtained.

%%%%%%%%%%%%%%
\subsection{Gravitational redshift}

\begin{figure}[h]
\centering
	\includegraphics[width=8cm]{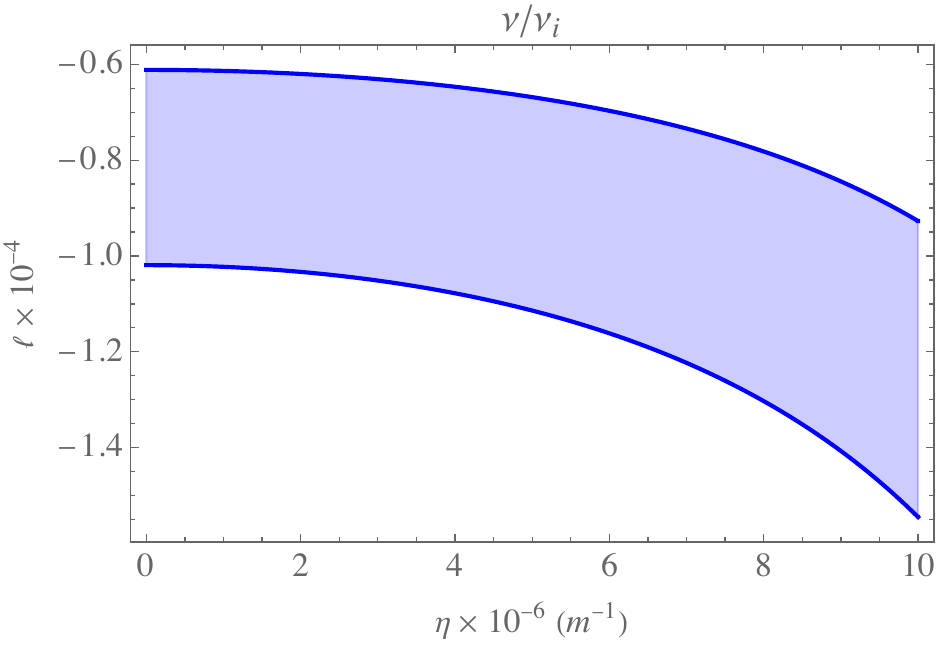} (a)\quad
    \includegraphics[width=8cm]{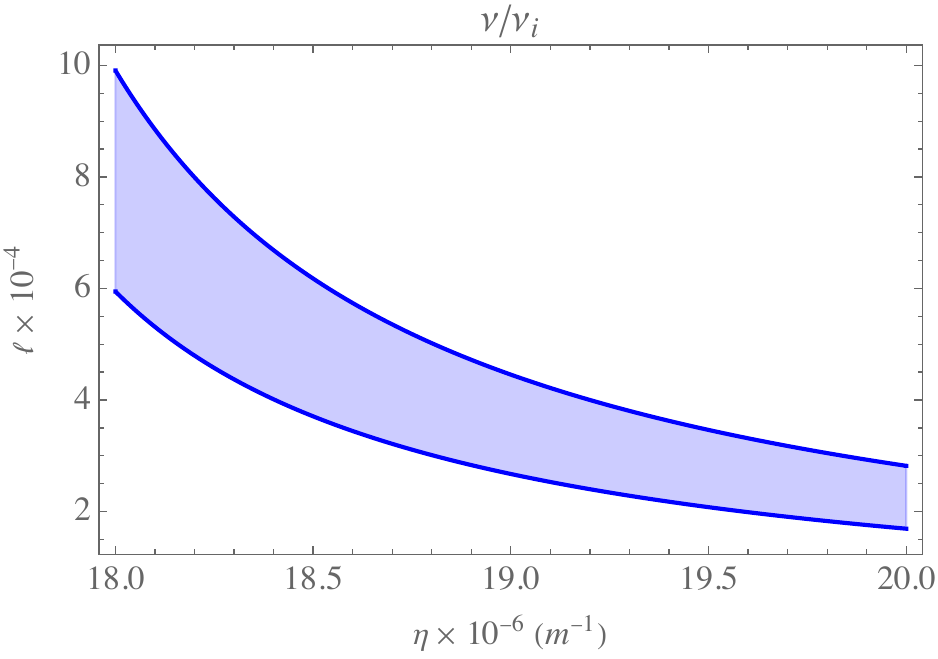} (b)
	\caption{The confidence intervals for $\ell$ and $\eta$ derived from the GP-A redshift measurements (specific parameter values are provided in Ref.~\cite{Vessot:1980zz}).}
	\label{fig:redshiftconst}
\end{figure}

The frequency shift experienced by photons as they traverse a static gravitational source can be derived using the relation \cite{Ryder:2009}
\begin{equation}\label{gredsh}
\frac{\nu}{\nu_i}=\sqrt{\frac{A(r)}{A(r_i)}},
\end{equation}
which applies to the spacetime metric given in Eq.~\eqref{eq:metric_0}. This equation arises from the presence of a time-like Killing vector field that characterizes the spacetime geometry. Here, $(r_i, \nu_i)$ and $(r, \nu)$ correspond to the source's initial and the observer's radial distance and frequency, respectively.
For scenarios near Earth's surface, the conditions $\ell \ll 1$ and $\frac{1}{2} \eta^2 \ell M r \ll 2M/r$ are satisfied. Under these approximations, Eq.~\eqref{gredsh} can be expanded as
\begin{equation}\label{gredsh2}
\frac{\nu}{\nu_i} \simeq \left(\frac{\nu}{\nu_i}\right)_{\mathrm{gr}} - 
\left(\frac{r - r_i}{r_i r}\right) M \ell + \frac{1}{4} \eta^2 \ell M (r - r_i),
\end{equation}
where 
\begin{equation}
\label{gredsh3}
\left(\frac{\nu}{\nu_i}\right)_{\mathrm{gr}} \equiv 1 - \frac{M}{r} + \frac{M}{r_i},
\end{equation}
represents the frequency shift due to general relativistic effects induced by the gravitational field of the source. This relation has been verified through the Gravity Probe A (GP-A) redshift experiment using a hydrogen maser, achieving an experimental accuracy of approximately $10^{-14}$ \cite{Vessot:1980zz}. As a result, the following inequality must hold:
\begin{equation}
\label{gredsh4} 
\left|
\left(\frac{r - r_i}{r_i r}\right) M \ell - \frac{1}{4} \eta^2 \ell M (r - r_i) \right| \lesssim 10^{-14}.
\end{equation}
Assuming the source is located at $r_i = r_\oplus$, corresponding to Earth's radius and mass $M = M_\oplus = 4.453 \times 10^{-3}$ m, and the observer is positioned on a satellite orbiting 15000 km above Earth's surface, Eq.~\eqref{gredsh4} imposes the constraint:
\begin{equation}
\label{gredsh5}
\left|4.877 \ell - 3.75 \ell M \eta^2 \right| \lesssim 1.
\end{equation}
This restriction yields approximate bounds of $|\ell| \sim 10^{-4}$ and $\eta \sim 10^{-6} \, \mathrm{m}^{-1}$ (see Fig.~\ref{fig:redshiftconst}).
%

%%%%%%%%%%%%%%%%%%
\subsection{Deflection of light}

\begin{figure}[h]
    \centering
    \includegraphics[width=8cm]{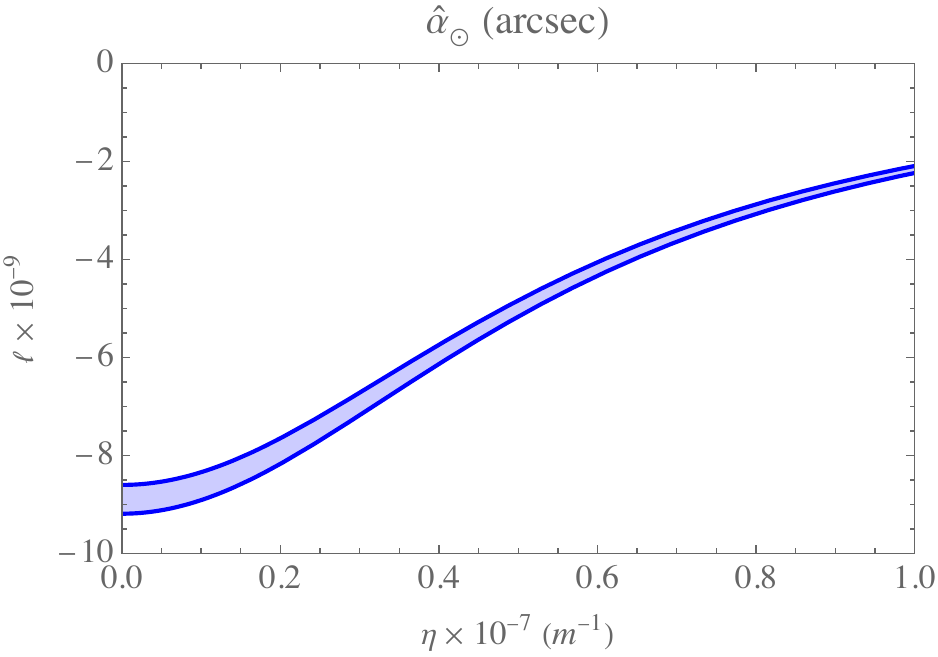}
    \caption{Parameter constraints on $\ell$ and $\eta$ based on solar deflection angle measurements.}
    \label{fig:lensigSun}
\end{figure}

In this section, we adopt the approach outlined in Ref.~\cite{takizawa_gravitational_2020} to compute the deflection angle of light passing near the Sun. This methodology leverages the Gauss-Bonnet theorem (GBT) and has been extended to accommodate non-asymptotically flat spacetimes.
Introducing $\dot{x}^\mu \equiv dx^\mu/ds$, we arrive at the following equation:
\begin{equation}\label{eq:geodesic_0}
\epsilon = -\frac{E^2}{A(r)} \dot{t}^2 + \frac{\dot{r}^2}{A(r)} + \frac{L^2}{r^2},
\end{equation}
where $E \equiv A(r) \dot{t}$ and $L \equiv r^2 \dot{\phi}$ are constants of motion. Consistent with prior discussions, equatorial plane motion ($\theta = \pi/2$) is assumed. The parameter $\epsilon$ characterizes the nature of the geodesics, with $\epsilon = 0$ representing null trajectories and $\epsilon = -1$ corresponding to time-like ones. For light rays (photons) traversing the black hole, the first-order angular equation of motion becomes \cite{Virbhadra:1999nm}
\begin{equation}
\left(\frac{\dot{r}}{\dot{\phi}}\right)^2 = \left(\frac{dr}{d\phi}\right)^2 = \frac{r^4}{b^2} - (1+\ell)r^2 + 2Mr - \frac{1}{2}\eta^2 M \ell r^3,
\label{eq:geoEq_0}
\end{equation}
where $b \equiv L/E$ denotes the impact parameter. Substituting $r = 1/u$, the equation transforms into
\begin{equation}\label{ue}
\left(\frac{du}{d\phi}\right)^2 = \frac{1}{b^2} - (1+\ell)u^2 + 2Mu^3 - \frac{1}{2}\eta^2 M \ell u,
\end{equation}
which reduces to the standard Schwarzschild case as $\ell \to 0$. Differentiating Eq.~\eqref{ue} with respect to $\phi$ yields
\begin{equation}
u'' + u = 3Mu^2 - \ell u - \frac{1}{4}\eta^2 M \ell,
\end{equation}
where primes denote derivatives with respect to $\phi$. Employing an iterative solution method, we find
\begin{equation}
u(\phi) \approx \frac{\sin\phi}{b} + \frac{M}{b^2}(1 + \cos^2\phi) - \frac{1}{4}M\ell\eta^2 + \mathcal{O}(\ell^2).
\label{eq:uphi_1}
\end{equation}
For the optical metric $\gamma_{ij} = \mathrm{diag}(A(r)^{-2}, r^2 A(r)^{-1})$, with $i, j = r, \phi$, the area element is $d\mathcal{S} = \sqrt{\mathrm{det}(\gamma_{ij})}\,dr d\phi = r A(r)^{-3/2} dr d\phi$. The Gaussian curvature $K$ on this surface, involving the lens, source, and observer, can then be expressed as
\begin{equation}
K = \frac{R_{r\phi r\phi}}{\gamma_{ij}},
\label{eq:K_0}
\end{equation}
and under the weak lensing approximation $M \ll b \ll (r_O, r_S)$ and $(r_O, r_S) \ll (\frac{1}{2}\eta^2 M \ell)^{-1}$, the curvature simplifies to
\begin{equation}
K \approx -\frac{2M}{r^3} - \frac{3}{2}\frac{M^2\ell\eta^2}{r^2} + \mathcal{O}(\ell^2).
\label{eq:K_app}
\end{equation}
Using the formulation of Ref.~\cite{Gibbons:2008rj,Ishihara:2016vdc,Li:2020wvn,takizawa_gravitational_2020}, the deflection angle is given as
\begin{equation}
\hat{\alpha} = \iint_{D_O+D_S} K\,d\mathcal{S} + \int_{P_O}^{P_S} \kappa_g\,d\mathfrak{l} + \phi_{OS},
\label{eq:alphahat_0}
\end{equation}
where $D_O$ and $D_S$ denote the observer and source regions, containing $P_O$ and $P_S$, respectively. Here, $\kappa_g$ is the geodesic curvature along the boundary and $\mathfrak{l}$ is the line element on this boundary. Substituting Eq.~\eqref{eq:uphi_1} and simplifying, we derive
\begin{equation}
\hat{\alpha} \approx \frac{2M}{b}\left(\sqrt{1 - b^2 u_S^2} + \sqrt{1 - b^2 u_O^2}\right) + \frac{1}{2}M^2\ell\eta^2\left(\frac{b u_S}{\sqrt{1 - b^2 u_S^2}} + \frac{b u_O}{\sqrt{1 - b^2 u_O^2}}\right) + \mathcal{O}(\ell^2).
\label{eq:alphahat_1}
\end{equation}
For large $r_{O, S}$, the $\ell\eta^2$ term vanishes, recovering the standard weak deflection angle $\hat{\alpha} \approx 4M/b$ for the Schwarzschild black hole. Observations near the Sun confirm $\hat{\alpha}_\odot \approx 1.7520$ arcsec (prograde) and $\hat{\alpha}_\odot \approx 1.7519$ arcsec (retrograde)~\cite{roy_study_2019}, leading to parameter constraints $\ell \sim -10^{-9}$ and $\eta \sim 10^{-7}~\mathrm{m}^{-1}$ (see Fig.~\ref{fig:lensigSun}).
%

%%%%%%%%%%%%%%%%%
\subsection{Gravitational time delay}

The Shapiro time delay, often referred to as the fourth test of general relativity, represents an intriguing phenomenon with significant observational relevance. This effect, describing the delay in radar echoes of electromagnetic signals traveling near massive objects, was experimentally verified around the time it was first proposed \cite{shapiro_fourth_1964,shapiro_fourth_1968,Virbhadra:2007kw}. Additionally, recent astrophysical observations indicate that this delay also occurs for two other massless energy carriers—neutrinos and gravitational waves—providing further support for the presence of dark matter \cite{boran_gw170817_2018}. 

In this subsection, we aim to calculate the Shapiro effect for photons passing near a black hole. Specifically, we derive the time difference between the emission and observation of a light ray sent from the point $P_1=(t_1, r_1)$, traveling to $P_2=(t_2,r_2)$, and returning to $P_1$. The total time interval can be expressed as
\begin{equation}
t_{12}=2\, t(r_1,r_0)+2\, t(r_2,r_0),
\end{equation}
where $r_0$ represents the point of closest approach to the central mass. Utilizing the definitions from the previous subsection, the radial velocity can be expressed as
\begin{equation}
\dot{r}=\dot{t}\frac{d r}{d t}=\frac{E}{A(r)}\frac{d r}{d t},
\end{equation}
which allows Eq.~\eqref{eq:geodesic_0} to be rewritten as:
\begin{equation}\label{ct}
\frac{E}{A(r)}\frac{d r}{d t}=\sqrt{E^2-\frac{L^2}{r^2}A(r)},
\end{equation}
for massless particles. At $r=r_0$, where the radial velocity vanishes, one finds that $b^{-2} = A(r_0)/r_0^2$. Thus, the coordinate time is given by
\begin{equation}
t(r,r_0)=\int_{r_0}^r \frac{d r}{A(r)\sqrt{1-\frac{r_0^2}{A(r_0)}\frac{A(r)}{r^2}}},
\end{equation}
for the segment of the journey between $r_0$ and $r$. To first-order corrections, this reduces to
\begin{equation}
t(r,r_0)\approx\sqrt{r^2-r_0^2}+t_M(r,r_0)+t_{\ell}(r,r_0)+t_{\eta}(r,r_0),
\end{equation}
where:
\begin{subequations}
	\begin{align}
	& t_M(r,r_0)=M\left[ \sqrt{\frac{r-r_0}{r+r_0}}+2\ln\left(\frac{r+\sqrt{r^2-r_0^2}}{r_0}\right)\right],\label{eq:tM}\\
	& t_{\ell}(r,r_0)=-\ell\sqrt{r^2-r_0^2},\label{eq:talpha}\\
	& t_{\eta}(r,r_0)=-\frac{1}{2}\ell M\eta^2 r_0^2\left[\sqrt{\frac{r-r_0}{r+r_0}}
	-\ln\left(\frac{r+\sqrt{r^2-r_0^2}}{r_0}\right)\right]
	-\frac{1}{4}\ell M\eta^2\left[r\,\sqrt{r^2-r_0^2}
	+r_0^2  \ln\left(\frac{r+\sqrt{r^2-r_0^2}}{r_0}\right)\right].\label{eq:tgamma}
	\end{align}
\end{subequations}
The time delay $\Delta t$ between the journey $P_1 \to P_2 \to P_1$ is defined as $\Delta t := t_{12} - t_{12}^{E}$, where $t_{12}^{E} = 2\left(\sqrt{r_1^2-r_0^2}+\sqrt{r_2^2-r_0^2}\right)$ is the travel time in Euclidean space. Substituting these expressions, we find
\begin{equation}\label{eq:Shapiro_0}
\Delta t =\Delta t_M+\Delta t_{\ell}+\Delta t_{\eta},
\end{equation}
where
\begin{subequations}
	\begin{align}
	& \Delta t_M = 2M\left[ \sqrt{\frac{r_1-r_0}{r_1+r_0}}+\sqrt{\frac{r_2-r_0}{r_2+r_0}}+2 \ln\left(\frac{\tilde{\mathfrak{t}}_{12}^{E}}{r_0^2}\right)\right],\label{eq:DeltatM}\\
	& \Delta t_{\ell}= -\ell t_{12}^{E},\label{eq:Deltatalpha}\\
	& \Delta t_{\eta} = -\ell M\eta^2\,r_0^2\left[\sqrt{\frac{r_1-r_0}{r_1+r_0}}
	+\sqrt{\frac{r_2-r_0}{r_2+r_0}}-\ln\left(\frac{\tilde{\mathfrak{t}}_{12}^{E}}{r_0^2}\right)
	\right]-\frac{1}{2}\ell M\eta^2 \left[ r_1\sqrt{r_1^2-r_0^2} +r_2\sqrt{r_2^2-r_0^2} +r_0^2\, \ln\left(\frac{\tilde{\mathfrak{t}}_{12}^{E}}{r_0^2}\right)\right],\label{eq:Deltatgamma}
	\end{align}
\end{subequations}
and $\tilde{\mathfrak{t}}_{12}^{E} = \left(r_1+\sqrt{r_1^2-r_0^2}\right)\left(r_2+\sqrt{r_2^2-r_0^2}\right)$. 

\begin{figure}[h]
	\centering
	\includegraphics[width=8cm]{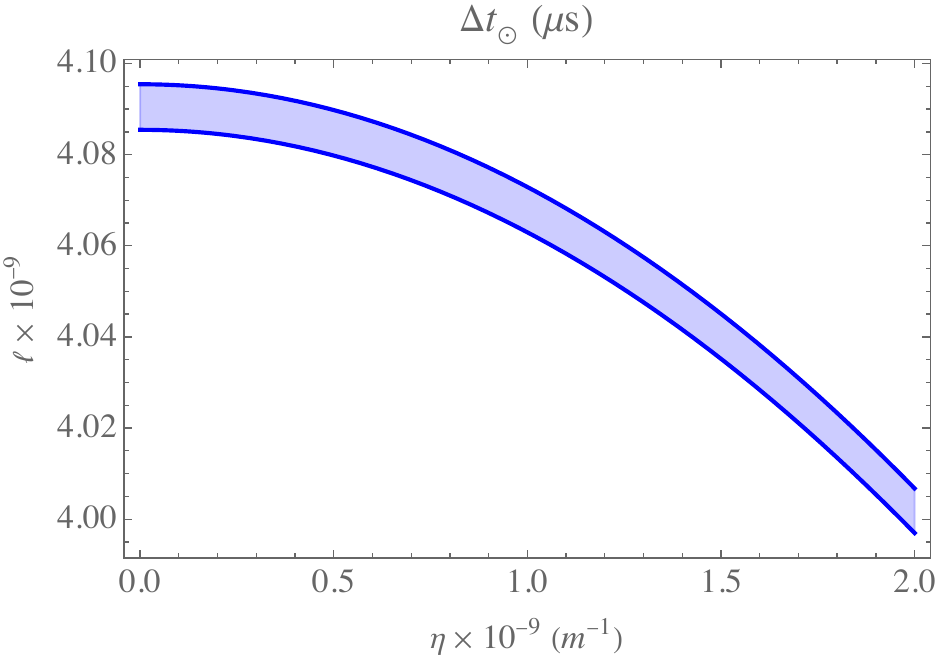}
	\caption{Constraints on $\ell$ and $\eta$ from solar system time delay observations.}
	\label{fig:ShapiroSun_1}
\end{figure}

Focusing on the solar system with $r_0\ll r_1, r_2$, the time delay approximates to
\begin{equation}
\Delta t_\odot \approx 4M\left[ 1+ \ln\left(\frac{4r_1r_2}{r_0^2}\right)\right]-2\ell\left(r_1+r_2\right)-\frac{1}{2}\ell M\eta^2\left[ r_1^2+r_2^2 -r_0^2 \ln\left(\frac{4r_1r_2}{r_0^2}\right)\right].
\label{eq:Shapiro_1}
\end{equation}
Substituting $M=M_{\odot}$ and $\ell= 0$, the classical Schwarzschild limit $\Delta t_{\mathrm{Sch}}=4M_{\odot}\left[ 1+ \ln\left(\frac{4r_1r_2}{r_0^2}\right)\right]$ is recovered. For typical distances, such as the Earth-Sun and Sun-Mars separations, and $r_0 \approx R_\odot + (5\times 10^6)$ m, $\Delta t_{\mathrm{Sch}}\approx 246\,\mathrm{\mu s}$. Measurements during the Viking mission reported an observational error of approximately $10\,\mathrm{ns}$ \cite{Viking:1979}, constraining $\ell\sim10^{-9}$ and $\eta\sim10^{-9}\,\mathrm{m}^{-1}$ (see Fig.~\ref{fig:ShapiroSun_1}).

Through these results, the positive domain of the Lorentz-violating parameter $\ell$ aligns with the observational data. While weak gravitational lensing favors negative $\ell$, the Shapiro delay supports positive $\ell$, yet both yield similar orders of magnitude for their constraints.

%
%\textcolor{Orange}{\textbf{Mohsen est\'a trabajando en esta parte!}$\longrightarrow \dots$}\\\\

%%%%%%%%%%%%%%Sect. V
\section{Conclusions}\label{sec:conclusions}

In this paper, we investigated the properties of a static, spherically symmetric black hole solution in a self-interacting KR field theory, minimally coupled with a global monopole constituent. The derived spacetime metric, characterized by the Lorentz-violating parameter $\ell$ and the monopole charge $\eta$, revealed several intriguing features, including deviations from Schwarzschild geometry and the emergence of effective topological and cosmological-like terms. 
Notably, the analysis of the causal structure revealed that for $\ell \leq 0$, the black hole always possesses an event horizon for any value of $\eta$, ensuring stability against horizon disappearance. However, for $\ell > 0$, horizons can merge, leading to the formation of extremal black holes, and eventually disappear, yielding naked singularities. This emphasizes the importance of $\ell$ in regulating the spacetime's physical structure. We then analyzed the thermodynamic behavior of the black hole for $\ell \leq 0$, observing that the contributions from the KR field and the monopole charge lead to modifications in the Hawking temperature, entropy, and Gibbs free energy. The interplay between $\ell$ and $\eta$ was found to significantly influence the thermodynamic stability. Specifically, larger values of $\eta$ enhance the Hawking temperature while decreasing the entropy for a fixed $\ell$, whereas $\ell$ impacts the temperature and entropy in opposite ways, depending on its sign. Our solar system tests provided observational constraints on $\ell$ and $\eta$. These constraints were derived from, the perihelion precession of planetary orbits, where $\ell$ and $\eta$ must satisfy the range $-0.0001 \leq \ell \leq 0.0001$ for $\eta \sim 10^{-6}\,\mathrm{m}^{-1}$, gravitational redshift data, which constrained $\ell$ to $\sim 10^{-4}$ and $\eta$ to $\sim 10^{-6}\,\mathrm{m}^{-1}$,  deflection of light near the Sun, with $\ell \sim -10^{-9}$ and $\eta \sim 10^{-7}\,\mathrm{m}^{-1}$, and the  Shapiro time delay measurements, which aligned $\ell$ to $\sim 10^{-9}$ and $\eta$ to $\sim 10^{-9}\,\mathrm{m}^{-1}$. The results consistently demonstrated that the Lorentz-violating parameter $\ell$ must be exceedingly small, aligning with existing constraints on deviations from Lorentz invariance. Meanwhile, the monopole charge $\eta$, though also tightly constrained, exhibited behavior that depended critically on its interplay with $\ell$. Our findings highlight the phenomenology introduced by the Lorentz-violating term and the monopole charge. The interplay of these parameters not only modifies the spacetime geometry but also imposes testable constraints, bridging theoretical predictions with astrophysical observations. In future work, we will focus on the constraints directly inferred from the black hole shadow and analyze the features imposed by incorporating an associated topological charge into the obtained black hole solution in the KR field. This study can deepen our understanding of the interplay between black hole geometry and the KR field's influence on observable phenomena.

%%%%%%%%%%%%%%%%%%%
\acknowledgments 

M.F. is supported by Universidad Central de Chile through project No. PDUCEN20240008. A.{\"O}. would like to acknowledge the contribution of the COST Action CA21106 - COSMIC WISPers in the Dark Universe: Theory, astrophysics and experiments (CosmicWISPers), the COST Action CA22113 - Fundamental challenges in theoretical physics (THEORY-CHALLENGES) and the COST Action CA23130 - Bridging high and low energies in search of quantum gravity (BridgeQG). We also thank EMU, TUBITAK and SCOAP3 for their support.

%%%%%%%%%%%%%%%%%%%
\section*{Data Availability Statement}

There are no data associated with the manuscript.

%\section{References}
%%%%%%%%%%%%%References
\bibliographystyle{ieeetr}
\bibliography{biblio_v1}

\begin{thebibliography}{100}

\bibitem{Kostelecky1989a}
V.~Kostelecky and S.~Samuel, ``Spontaneous breaking of lorentz symmetry in string theory,'' {\em Phys. Rev. D}, vol.~39, p.~683, 1989.

\bibitem{Alfaro2002}
J.~Alfaro, H.~Morales-Tecotl, and L.~Urrutia, ``Loop quantum gravity and light propagation,'' {\em Phys. Rev. D}, vol.~65, p.~103509, 2002.

\bibitem{Horava2009a}
P.~Horava, ``Quantum gravity at a lifshitz point,'' {\em Phys. Rev. D}, vol.~79, p.~084008, 2009.

\bibitem{Carroll2001}
S.~Carroll, J.~Harvey, V.~Kostelecky, C.~Lane, and T.~Okamoto, ``Noncommutative field theory and lorentz violation,'' {\em Phys. Rev. Lett.}, vol.~87, p.~141601, 2001.

\bibitem{Jacobson2001}
T.~Jacobson and D.~Mattingly, ``Gravity with a dynamical preferred frame,'' {\em Phys. Rev. D}, vol.~64, p.~024028, 2001.

\bibitem{Dubovsky2005}
S.~Dubovsky, P.~Tinyakov, and I.~Tkachev, ``Massive graviton as a testable cold dark matter candidate,'' {\em Phys. Rev. Lett.}, vol.~94, p.~181102, 2005.

\bibitem{Bengochea2009}
G.~Bengochea and R.~Ferraro, ``Dark torsion as the cosmic speed-up,'' {\em Phys. Rev. D}, vol.~79, p.~124019, 2009.

\bibitem{Cohen2006}
A.~Cohen and S.~Glashow, ``Very special relativity,'' {\em Phys. Rev. Lett.}, vol.~97, p.~021601, 2006.

\bibitem{Lehnert:2006mn}
R.~Lehnert, ``{CPT and Lorentz-symmetry breaking: A Review},'' {\em Frascati Phys. Ser.}, vol.~43, pp.~131--154, 2007.

\bibitem{Kostelecky:2021bsb}
V.~A. Kosteleck\'y, R.~Lehnert, N.~McGinnis, M.~Schreck, and B.~Seradjeh, ``{Lorentz violation in Dirac and Weyl semimetals},'' {\em Phys. Rev. Res.}, vol.~4, no.~2, p.~023106, 2022.

\bibitem{Lehnert:2004be}
R.~Lehnert and R.~Potting, ``{The Cerenkov effect in Lorentz-violating vacua},'' {\em Phys. Rev. D}, vol.~70, p.~125010, 2004.
\newblock [Erratum: Phys.Rev.D 70, 129906 (2004)].

\bibitem{Diaz:2013saa}
J.~S. D\'\i{}az, A.~Kosteleck\'y, and R.~Lehnert, ``{Relativity violations and beta decay},'' {\em Phys. Rev. D}, vol.~88, no.~7, p.~071902, 2013.

\bibitem{Cambiaso:2012vb}
M.~Cambiaso, R.~Lehnert, and R.~Potting, ``{Massive photons and Lorentz violation},'' {\em Phys. Rev. D}, vol.~85, p.~085023, 2012.

\bibitem{Seifert:2010uu}
M.~D. Seifert, ``{A Monopole solution in a Lorentz-violating field theory},'' {\em Phys. Rev. Lett.}, vol.~105, p.~201601, 2010.

\bibitem{Gullu:2020qzu}
I.~G\"ull\"u and A.~\"Ovg\"un, ``{Schwarzschild-like black hole with a topological defect in bumblebee gravity},'' {\em Annals Phys.}, vol.~436, p.~168721, 2022.

\bibitem{Ovgun:2018xys}
A.~\"Ovg\"un, K.~Jusufi, and I.~Sakall\i{}, ``{Exact traversable wormhole solution in bumblebee gravity},'' {\em Phys. Rev. D}, vol.~99, no.~2, p.~024042, 2019.

\bibitem{Oliveira:2018oha}
R.~Oliveira, D.~M. Dantas, V.~Santos, and C.~A.~S. Almeida, ``{Quasinormal modes of bumblebee wormhole},'' {\em Class. Quant. Grav.}, vol.~36, no.~10, p.~105013, 2019.

\bibitem{Lambiase:2024uzy}
G.~Lambiase, R.~C. Pantig, and A.~\"Ovg\"un, ``{Weak field deflection angle and analytical parameter estimation of the Lorentz-violating Bumblebee parameter through the black hole shadow using EHT data},'' {\em EPL}, vol.~148, no.~4, p.~49001, 2024.

\bibitem{Carleo:2022qlv}
A.~Carleo, G.~Lambiase, and L.~Mastrototaro, ``{Energy extraction via magnetic reconnection in Lorentz breaking Kerr\textendash{}Sen and Kiselev black holes},'' {\em Eur. Phys. J. C}, vol.~82, no.~9, p.~776, 2022.

\bibitem{Lambiase:2017adh}
G.~Lambiase and F.~Scardigli, ``{Lorentz violation and generalized uncertainty principle},'' {\em Phys. Rev. D}, vol.~97, no.~7, p.~075003, 2018.

\bibitem{AraujoFilho:2024ykw}
A.~A. Ara\'ujo~Filho, J.~R. Nascimento, A.~Y. Petrov, and P.~J. Porf\'\i{}rio, ``{An exact stationary axisymmetric vacuum solution within a metric-affine bumblebee gravity},'' {\em JCAP}, vol.~07, p.~004, 2024.

\bibitem{Filho:2022yrk}
A.~A.~A. Filho, J.~R. Nascimento, A.~Y. Petrov, and P.~J. Porf\'\i{}rio, ``{Vacuum solution within a metric-affine bumblebee gravity},'' {\em Phys. Rev. D}, vol.~108, no.~8, p.~085010, 2023.

\bibitem{Araujo:2024tiy}
R.~Araujo, T.~Mariz, J.~R. Nascimento, and A.~Y. Petrov, ``{Derivative four-fermion model, effective action and bumblebee generation},'' {\em Eur. Phys. J. C}, vol.~84, no.~10, p.~1034, 2024.

\bibitem{Nascimento:2023auz}
J.~R. Nascimento, G.~J. Olmo, A.~Y. Petrov, and P.~J. Porfirio, ``{On metric-affine bumblebee model coupled to scalar matter},'' {\em Nucl. Phys. B}, vol.~1004, p.~116577, 2024.

\bibitem{Heidari:2024bvd}
N.~Heidari, C.~F.~B. Macedo, A.~A.~A. Filho, and H.~Hassanabadi, ``{Scattering effects of bumblebee gravity in metric-affine formalism},'' {\em Eur. Phys. J. C}, vol.~84, no.~11, p.~1221, 2024.

\bibitem{Hosseinifar:2024wwe}
F.~Hosseinifar, A.~A.~A. Filho, M.~Y. Zhang, H.~Chen, and H.~Hassanabadi, ``{Shadows, greybody factors, emission rate, topological charge, and phase transitions for a charged black hole with a Kalb-Ramond field background},'' 7 2024.

\bibitem{AraujoFilho:2024ctw}
A.~A. Ara\'ujo~Filho, ``{Particle creation and evaporation in Kalb-Ramond gravity},'' 11 2024.

\bibitem{Maluf:2018jwc}
R.~V. Maluf, A.~A. Ara\'ujo~Filho, W.~T. Cruz, and C.~A.~S. Almeida, ``{Antisymmetric tensor propagator with spontaneous Lorentz violation},'' {\em EPL}, vol.~124, no.~6, p.~61001, 2018.

\bibitem{Kostelecky2004a}
V.~Kostelecky, ``Gravity, lorentz violation, and the standard model,'' {\em Phys. Rev. D}, vol.~69, p.~105009, 2004.

\bibitem{Kostelecky1989}
V.~Kostelecky and S.~Samuel, ``Gravitational phenomenology in higher dimensional theories and strings,'' {\em Phys. Rev. D}, vol.~40, p.~1886, 1989.

\bibitem{Kostelecky1989b}
V.~Kostelecky and S.~Samuel, ``Phenomenological gravitational constraints on strings and higher dimensional theories,'' {\em Phys. Rev. Lett.}, vol.~63, p.~224, 1989.

\bibitem{Bailey2006}
Q.~Bailey and V.~Kostelecky, ``Signals for lorentz violation in post-newtonian gravity,'' {\em Phys. Rev. D}, vol.~74, p.~045001, 2006.

\bibitem{Bluhm2008a}
R.~Bluhm, N.~Gagne, R.~Potting, and A.~Vrublevskis, ``Constraints and stability in vector theories with spontaneous lorentz violation,'' {\em Phys. Rev. D}, vol.~77, p.~125007, 2008.

\bibitem{Casana2018}
R.~Casana, A.~Cavalcante, F.~P. Poulis, and E.~B. Santos, ``Exact schwarzschild-like solution in a bumblebee gravity model,'' {\em Phys. Rev. D}, vol.~97, p.~104001, 2018.

\bibitem{Ovgun2018}
A.~\"Ovg\"un, K.~Jusufi, and I.~Sakalli, ``Gravitational lensing under the effect of weyl and bumblebee gravities: Applications of gauss-bonnet theorem,'' {\em Annals Phys.}, vol.~399, p.~193, 2018.

\bibitem{Lambiase:2023zeo}
G.~Lambiase, L.~Mastrototaro, R.~C. Pantig, and A.~Ovgun, ``{Probing Schwarzschild-like black holes in metric-affine bumblebee gravity with accretion disk, deflection angle, greybody bounds, and neutrino propagation},'' {\em JCAP}, vol.~12, p.~026, 2023.

\bibitem{Kuang:2022xjp}
X.-M. Kuang and A.~\"Ovg\"un, ``{Strong gravitational lensing and shadow constraint from M87* of slowly rotating Kerr-like black hole},'' {\em Annals Phys.}, vol.~447, p.~169147, 2022.

\bibitem{Mangut:2023oxa}
M.~Mangut, H.~G\"ursel, S.~Kanzi, and I.~Sakall\i{}, ``{Probing the Lorentz Invariance Violation via Gravitational Lensing and Analytical Eigenmodes of Perturbed Slowly Rotating Bumblebee Black Holes},'' {\em Universe}, vol.~9, no.~5, p.~225, 2023.

\bibitem{Pantig:2024ixc}
R.~C. Pantig, S.~Kala, A.~\"Ovg\"un, and N.~J. L.~S. Lobos, ``{Testing black holes with cosmological constant in Einstein-bumblebee gravity through the black hole shadow using EHT data and deflection angle},'' 10 2024.

\bibitem{Panotopoulos:2024jtn}
G.~Panotopoulos and A.~\"Ovg\"un, ``{Strange Quark Stars and Condensate Dark Stars in Bumblebee Gravity},'' 9 2024.

\bibitem{Sakalli:2023pgn}
I.~Sakall\i{} and E.~Y\"or\"uk, ``{Modified Hawking radiation of Schwarzschild-like black hole in bumblebee gravity model},'' {\em Phys. Scripta}, vol.~98, no.~12, p.~125307, 2023.

\bibitem{Oliveira2021}
R.~Oliveira, D.~M. Dantas, and C.~A.~S. Almeida, ``Quasinormal frequencies for a black hole in a bumblebee gravity,'' {\em EPL}, vol.~135, p.~10003, 2021.

\bibitem{Maluf2021}
R.~V. Maluf and J.~C.~S. Neves, ``Black holes with a cosmological constant in bumblebee gravity,'' {\em Phys. Rev. D}, vol.~103, p.~044002, 2021.

\bibitem{Xu2023}
R.~Xu, D.~Liang, and L.~Shao, ``Static spherical vacuum solutions in the bumblebee gravity model,'' {\em Phys. Rev. D}, vol.~107, p.~024011, 2023.

\bibitem{Ding2020a}
C.~Ding, C.~Liu, R.~Casana, and A.~Cavalcante, ``Exact kerr-like solution and its shadow in a gravity model with spontaneous lorentz symmetry breaking,'' {\em Eur. Phys. J. C}, vol.~80, p.~178, 2020.

\bibitem{Ding2021a}
C.~Ding and X.~Chen, ``Slowly rotating einstein-bumblebee black hole solution and its greybody factor in a lorentz violation model,'' {\em Chin. Phys. C}, vol.~45, p.~025106, 2021.

\bibitem{Liang2022}
D.~Liang, R.~Xu, X.~Lu, and L.~Shao, ``Polarizations of gravitational waves in the bumblebee gravity model,'' {\em Phys. Rev. D}, vol.~106, p.~124019, 2022.

\bibitem{Amarilo2023}
K.~M. Amarilo, M.~B.~F. Filho, A.~A.~A. Filho, and J.~A. A.~S. Reis, ``Gravitational waves effects in a lorentz-violating scenario,''

\bibitem{altschul_lorentz_2010}
B.~Altschul, Q.~G. Bailey, and V.~A. Kostelecký, ``Lorentz violation with an antisymmetric tensor,'' {\em Physical Review D}, vol.~81, p.~065028, Mar. 2010.

\bibitem{Kalb1974}
M.~Kalb and P.~Ramond, ``Classical direct interstring action,'' {\em Phys. Rev. D}, vol.~9, p.~2273, 1974.

\bibitem{Kao1996}
W.~Kao, W.~Dai, S.-Y. Wang, T.-K. Chyi, and S.-Y. Lin, ``Induced einstein-kalb-ramond theory and the black hole,'' {\em Phys. Rev. D}, vol.~53, p.~2244, 1996.

\bibitem{Kar2003}
S.~Kar, S.~SenGupta, and S.~Sur, ``Static spherisymmetric solutions, gravitational lensing and perihelion precession in einstein-kalb-ramond theory,'' {\em Phys. Rev. D}, vol.~67, p.~044005, 2003.
\newblock arXiv:hep-th/0210176.

\bibitem{Chakraborty2017}
S.~Chakraborty and S.~SenGupta, ``Strong gravitational lensing — a probe for extra dimensions and kalb-ramond field,'' {\em JCAP}, vol.~07, p.~045, 2017.
\newblock arXiv:1611.06936.

\bibitem{Junior:2024vdk}
E.~L.~B. Junior, J.~T. S.~S. Junior, F.~S.~N. Lobo, M.~E. Rodrigues, D.~Rubiera-Garcia, L.~F.~D. da~Silva, and H.~A. Vieira, ``{Gravitational lensing of a Schwarzschild-like black hole in Kalb-Ramond gravity},'' {\em Phys. Rev. D}, vol.~110, no.~2, p.~024077, 2024.

\bibitem{Zahid:2024ohn}
M.~Zahid, J.~Rayimbaev, N.~Kurbonov, S.~Ahmedov, C.~Shen, and A.~Abdujabbarov, ``{Electric Penrose, circular orbits and collisions of charged particles near charged black holes in Kalb\textendash{}Ramond gravity},'' {\em Eur. Phys. J. C}, vol.~84, no.~7, p.~706, 2024.

\bibitem{Jumaniyozov:2024eah}
S.~Jumaniyozov, S.~U. Khan, J.~Rayimbaev, A.~Abdujabbarov, S.~Urinbaev, and S.~Murodov, ``{Circular motion and QPOs near black holes in Kalb\textendash{}Ramond gravity},'' {\em Eur. Phys. J. C}, vol.~84, no.~9, p.~964, 2024.

\bibitem{Ditta:2024lnb}
A.~Ditta, F.~Javed, A.~Bouzenada, G.~Mustafa, A.~Mahmood, F.~Atamurotov, and V.~Khamidov, ``{Thermal chemistry of Anti-de-Sitter black holes in Kalb-Ramond gravity},'' {\em JHEAp}, vol.~45, pp.~62--74, 2025.

\bibitem{al-Badawi:2024pdx}
A.~al~Badawi, S.~Shaymatov, and I.~Sakall\i{}, ``{Geodesics structure and deflection angle of electrically charged black holes in gravity with a background Kalb\textendash{}Ramond field},'' {\em Eur. Phys. J. C}, vol.~84, no.~8, p.~825, 2024.

\bibitem{Ortiqboev:2024mtk}
D.~Ortiqboev, F.~Javed, F.~Atamurotov, A.~Abdujabbarov, and G.~Mustafa, ``{Energy extraction and Keplerian fundamental frequencies in the Kalb\textendash{}Ramond gravity},'' {\em Phys. Dark Univ.}, vol.~46, p.~101615, 2024.

\bibitem{Junior:2024ety}
E.~L.~B. Junior, J.~T. S.~S. Junior, F.~S.~N. Lobo, M.~E. Rodrigues, D.~Rubiera-Garcia, L.~F.~D. da~Silva, and H.~A. Vieira, ``{Spontaneous Lorentz symmetry-breaking constraints in Kalb\textendash{}Ramond gravity},'' {\em Eur. Phys. J. C}, vol.~84, no.~12, p.~1257, 2024.

\bibitem{Ali:2023amn}
R.~Ali, R.~Babar, M.~Asgher, and G.~Mustafa, ``{Quantum Gravity Evolution of the Kalb\textendash{}Ramond like black hole},'' {\em Chin. J. Phys.}, vol.~86, pp.~269--279, 2023.

\bibitem{Al-Badawi:2023xig}
A.~Al-Badawi and A.~Kraishan, ``{Fermionic greybody factors and quasinormal modes of black holes in Kalb\textendash{}Ramond gravity},'' {\em Annals Phys.}, vol.~458, p.~169467, 2023.

\bibitem{Rahaman:2023swt}
F.~Rahaman, A.~Aziz, T.~Manna, A.~Islam, N.~A. Pundeer, and S.~Islam, ``{Deflection of massive body around wormholes in Einstein\textendash{}Kalb\textendash{}Ramond spacetime},'' {\em Phys. Dark Univ.}, vol.~42, p.~101287, 2023.

\bibitem{Baruah:2023rhd}
A.~Baruah, A.~\"Ovg\"un, and A.~Deshamukhya, ``{Quasinormal modes and bounding greybody factors of GUP-corrected black holes in Kalb\textendash{}Ramond gravity},'' {\em Annals Phys.}, vol.~455, p.~169393, 2023.

\bibitem{Nair2022}
K.~Nair and A.~Thomas, ``Kalb-ramond field-induced cosmological bounce in generalized teleparallel gravity,'' {\em Phys. Rev. D}, vol.~105, p.~103505, 2022.
\newblock arXiv:2112.11945.

\bibitem{Fu2012}
C.-E. Fu, Y.-X. Liu, K.~Yang, and S.-W. Wei, ``Q-form fields on p-branes,'' {\em JHEP}, vol.~10, p.~060, 2012.
\newblock arXiv:1207.3152.

\bibitem{Chakraborty2016}
S.~Chakraborty and S.~SenGupta, ``Solutions on a brane in a bulk spacetime with kalb-ramond field,'' {\em Annals Phys.}, vol.~367, p.~258, 2016.
\newblock arXiv:1412.7783.

\bibitem{Lessa2020}
L.~Lessa, J.~Silva, R.~Maluf, and C.~Almeida, ``Modified black hole solution with a background kalb-ramond field,'' {\em Eur. Phys. J. C}, vol.~80, p.~335, 2020.
\newblock arXiv:1911.10296.

\bibitem{Atamurotov2022}
F.~Atamurotov, D.~Ortiqboev, A.~Abdujabbarov, and G.~Mustafa, ``Particle dynamics and gravitational weak lensing around black hole in the kalb-ramond gravity,'' {\em Eur. Phys. J. C}, vol.~82, p.~659, 2022.

\bibitem{Kumar2020c}
R.~Kumar, S.~Ghosh, and A.~Wang, ``Gravitational deflection of light and shadow cast by rotating kalb-ramond black holes,'' {\em Phys. Rev. D}, vol.~101, p.~104001, 2020.
\newblock arXiv:2001.00460.

\bibitem{Lessa2021}
L.~Lessa, R.~Oliveira, J.~Silva, and C.~Almeida, ``Traversable wormhole solution with a background kalb-ramond field,'' {\em Annals Phys.}, vol.~433, p.~168604, 2021.
\newblock arXiv:2010.05298.

\bibitem{Maluf2022}
R.~Maluf and C.~Muniz, ``Exact solution for a traversable wormhole in a curvature-coupled antisymmetric background field,'' {\em Eur. Phys. J. C}, vol.~82, p.~445, 2022.
\newblock arXiv:2110.12202.

\bibitem{Maluf2022a}
R.~Maluf and J.~Neves, ``Bianchi type i cosmology with a kalb-ramond background field,'' {\em Eur. Phys. J. C}, vol.~82, p.~135, 2022.
\newblock arXiv:2111.13165.

\bibitem{yang_static_2023}
K.~Yang, Y.-Z. Chen, Z.-Q. Duan, and J.-Y. Zhao, ``Static and spherically symmetric black holes in gravity with a background {Kalb}-{Ramond} field,'' {\em Physical Review D}, vol.~108, p.~124004, Dec. 2023.

\bibitem{liu_static_2024}
W.~Liu, D.~Wu, and J.~Wang, ``Static neutral black holes in {Kalb}-{Ramond} gravity,'' {\em Journal of Cosmology and Astroparticle Physics}, vol.~2024, p.~017, Sept. 2024.

\bibitem{duan_electrically_2024}
Z.-Q. Duan, J.-Y. Zhao, and K.~Yang, ``Electrically charged black holes in gravity with a background {Kalb}–{Ramond} field,'' {\em The European Physical Journal C}, vol.~84, p.~798, Aug. 2024.

\bibitem{Vilenkin1981}
A.~Vilenkin, ``Gravitational field of vacuum domain walls and strings,'' {\em Phys. Rev. D}, vol.~23, pp.~852--857, 1981.

\bibitem{Vilenkin1982}
A.~Vilenkin, ``Cosmological string theories,'' {\em Nucl. Phys. B}, vol.~196, pp.~240--258, 1982.

\bibitem{barriola_gravitational_1989}
M.~Barriola and A.~Vilenkin, ``Gravitational field of a global monopole,'' {\em Physical Review Letters}, vol.~63, pp.~341--343, July 1989.

\bibitem{Dadhich:1997mh}
N.~Dadhich, K.~Narayan, and U.~A. Yajnik, ``{Schwarzschild black hole with global monopole charge},'' {\em Pramana}, vol.~50, pp.~307--314, 1998.

\bibitem{Bronnikov}
K.~Bronnikov, B.~Meierovich, and E.~Podolyak, ``Black holes with a nonlinear electromagnetic field,'' {\em J. Exp. Theor. Phys.}, vol.~95, p.~392, 2002.

\bibitem{Preskill}
J.~Preskill, ``Gauge-invariant extension of the quantum chromodynamics,'' {\em Phys. Rev. Lett.}, vol.~43, p.~1365, 1979.

\bibitem{Barriola}
M.~Barriola and A.~Vilenkin, ``Gravitational monopoles and solitons,'' {\em Phys. Rev. Lett.}, vol.~63, p.~341, 1989.

\bibitem{lessa_modified_2020}
L.~A. Lessa, J.~E.~G. Silva, R.~V. Maluf, and C.~A.~S. Almeida, ``Modified black hole solution with a background {Kalb}–{Ramond} field,'' {\em The European Physical Journal C}, vol.~80, p.~335, Apr. 2020.

\bibitem{bluhm_spontaneous_2008}
R.~Bluhm, S.-H. Fung, and V.~A. Kostelecký, ``Spontaneous {Lorentz} and diffeomorphism violation, massive modes, and gravity,'' {\em Physical Review D}, vol.~77, p.~065020, Mar. 2008.

\bibitem{fathi_study_2022}
M.~Fathi, M.~Olivares, and J.~R. Villanueva, ``Study of null and time-like geodesics in the exterior of a {Schwarzschild} black hole with quintessence and cloud of strings,'' {\em The European Physical Journal C}, vol.~82, p.~629, July 2022.

\bibitem{mannheim_exact_1989}
P.~D. Mannheim and D.~Kazanas, ``Exact vacuum solution to conformal {Weyl} gravity and galactic rotation curves,'' {\em The Astrophysical Journal}, vol.~342, p.~635, July 1989.

\bibitem{bardeen_four_1973}
J.~M. Bardeen, B.~Carter, and S.~W. Hawking, ``The four laws of black hole mechanics,'' {\em Communications in Mathematical Physics}, vol.~31, pp.~161--170, June 1973.

\bibitem{kastor_enthalpy_2009}
D.~Kastor, S.~Ray, and J.~Traschen, ``Enthalpy and the mechanics of {AdS} black holes,'' {\em Classical and Quantum Gravity}, vol.~26, p.~195011, Oct. 2009.

\bibitem{wei_insight_2015}
S.-W. Wei and Y.-X. Liu, ``Insight into the {Microscopic} {Structure} of an {AdS} {Black} {Hole} from a {Thermodynamical} {Phase} {Transition},'' {\em Physical Review Letters}, vol.~115, p.~111302, Sept. 2015.

\bibitem{kubiznak_black_2017}
D.~Kubizňák, R.~B. Mann, and M.~Teo, ``Black hole chemistry: thermodynamics with {Lambda},'' {\em Classical and Quantum Gravity}, vol.~34, p.~063001, Mar. 2017.

\bibitem{yang_kinetics_2022}
S.-J. Yang, R.~Zhou, S.-W. Wei, and Y.-X. Liu, ``Kinetics of a phase transition for a {Kerr}-{AdS} black hole on the free-energy landscape,'' {\em Physical Review D}, vol.~105, p.~084030, Apr. 2022.

\bibitem{mbarek_reverse_2019}
S.~Mbarek and R.~B. Mann, ``Reverse {Hawking}-{Page} phase transition in de {Sitter} black holes,'' {\em Journal of High Energy Physics}, vol.~2019, p.~103, Feb. 2019.

\bibitem{cornbleet93}
S.~Cornbleet, ``Elementary derivation of the advance of the perihelion of a planetary orbit,'' {\em American Journal of Physics}, vol.~61, no.~7, pp.~650--651, 1993.

\bibitem{Ryder:2009}
L.~Ryder, {\em Introduction to General Relativity}.
\newblock Cambridge University Press, 2009.

\bibitem{Vessot:1980zz}
R.~F.~C. Vessot {\em et~al.}, ``{Test of Relativistic Gravitation with a Space-Borne Hydrogen Maser},'' {\em Phys. Rev. Lett.}, vol.~45, pp.~2081--2084, 1980.

\bibitem{takizawa_gravitational_2020}
K.~Takizawa, T.~Ono, and H.~Asada, ``Gravitational deflection angle of light: {Definition} by an observer and its application to an asymptotically nonflat spacetime,'' {\em Physical Review D}, vol.~101, p.~104032, May 2020.

\bibitem{Virbhadra:1999nm}
K.~S. Virbhadra and G.~F.~R. Ellis, ``{Schwarzschild black hole lensing},'' {\em Phys. Rev. D}, vol.~62, p.~084003, 2000.

\bibitem{Gibbons:2008rj}
G.~W. Gibbons and M.~C. Werner, ``{Applications of the Gauss-Bonnet theorem to gravitational lensing},'' {\em Class. Quant. Grav.}, vol.~25, p.~235009, 2008.

\bibitem{Ishihara:2016vdc}
A.~Ishihara, Y.~Suzuki, T.~Ono, T.~Kitamura, and H.~Asada, ``{Gravitational bending angle of light for finite distance and the Gauss-Bonnet theorem},'' {\em Phys. Rev. D}, vol.~94, no.~8, p.~084015, 2016.

\bibitem{Li:2020wvn}
Z.~Li, G.~Zhang, and A.~\"Ovg\"un, ``{Circular Orbit of a Particle and Weak Gravitational Lensing},'' {\em Phys. Rev. D}, vol.~101, no.~12, p.~124058, 2020.

\bibitem{roy_study_2019}
S.~Roy and A.~K. Sen, ``Study of gravitational deflection of light ray,'' {\em Journal of Physics: Conference Series}, vol.~1330, p.~012002, Oct. 2019.

\bibitem{shapiro_fourth_1964}
I.~I. Shapiro, ``Fourth {Test} of {General} {Relativity},'' {\em Physical Review Letters}, vol.~13, pp.~789--791, Dec. 1964.

\bibitem{shapiro_fourth_1968}
I.~I. Shapiro, G.~H. Pettengill, M.~E. Ash, M.~L. Stone, W.~B. Smith, R.~P. Ingalls, and R.~A. Brockelman, ``Fourth {Test} of {General} {Relativity}: {Preliminary} {Results},'' {\em Physical Review Letters}, vol.~20, pp.~1265--1269, May 1968.

\bibitem{Virbhadra:2007kw}
K.~S. Virbhadra and C.~R. Keeton, ``{Time delay and magnification centroid due to gravitational lensing by black holes and naked singularities},'' {\em Phys. Rev. D}, vol.~77, p.~124014, 2008.

\bibitem{boran_gw170817_2018}
S.~Boran, S.~Desai, E.~Kahya, and R.~Woodard, ``{GW170817} falsifies dark matter emulators,'' {\em Physical Review D}, vol.~97, p.~041501, Feb. 2018.

\bibitem{Viking:1979}
R.~D. Reasenberg, I.~I. Shapiro, P.~E. MacNeil, R.~B. Goldstein, J.~C. Breidenthal, J.~P. Brenkle, D.~L. Cain, T.~M. Kaufman, T.~A. Komarek, and A.~I. Zygielbaum, ``Viking relativity experiment - {Verification} of signal retardation by solar gravity,'' {\em The Astrophysical Journal}, vol.~234, p.~L219, Dec. 1979.

\end{thebibliography}
%\bibliography{DymnikovaTestEnero2024}

%\bibliography{references}
%\bibliographystyle{apsrev}
%\bibliographystyle{aasjournal}
\end{document}